\documentclass[useAMS,usenatbib]{mn2e}
\usepackage{journals}
\usepackage[latin1]{inputenc}
\usepackage[T1]{fontenc}
\usepackage{graphics}         
\usepackage{amsfonts}
\usepackage{amsmath}
\usepackage{caption}
\usepackage{natbib}
\usepackage{multicol}
\usepackage{layout}
\usepackage{amssymb}
\usepackage[pdftex]{graphicx,color}
\usepackage[draft,colorlinks=true,pdfstartview=FitV,linkcolor=red,citecolor=blue,urlcolor=magenta]{hyperref}

\def\simlt{\lower.5ex\hbox{$\; \buildrel < \over \sim \;$}}
\def\simgt{\lower.5ex\hbox{$\; \buildrel > \over \sim \;$}}
\def\simpt{\lower.5ex\hbox{$\; \buildrel \propto \over \sim \;$}}

\title[Weak Lensing Statistics]
	{Disentangling dark sector models using 
          weak lensing statistics}
		\author[Giocoli et al. 2015] 
        	{\parbox{\textwidth}{
        	Carlo Giocoli$^{1,2,3,4}$\thanks{E-mail:
             	\href{mailto:carlo.giocoli@lam.fr}{carlo.giocoli@lam.fr}}, 
		R. Benton Metcalf$^2$,
		Marco Baldi$^{2,3,4}$,
		Massimo Meneghetti$^{3,4,5}$,
     	        Lauro Moscardini$^{2,3,4}$,
	        Margarita Petkova$^6$
				} \\ \\               
            $^{1}$ Aix Marseille Universit\'{e}, CNRS, LAM (Laboratoire d'Astrophysique de Marseille) UMR 7326, 13388 Marseille, France \\
            $^{2}$ Dipartimento di Fisica e Astronomia, Alma Mater Studiorum Universit\`{a} di 
                   Bologna, viale Berti Pichat, 6/2, 40127 Bologna, Italy\\ 
            $^{3}$ INAF - Osservatorio Astronomico di
                   Bologna, via Ranzani 1, 40127 Bologna, Italy \\ 
            $^{4}$ INFN - Sezione di Bologna, viale Berti Pichat 6/2, 
                   40127 Bologna, Italy \\ 
            $^{5}$ Jet Propulsion Laboratory, 4800 Oak Grove Dr. Pasadena, CA 91109, USA \\ 
            $^{6}$ Department of Physics, Ludwig-Maximilians-Universitaet, 
                   Scheinerstr. 1, D-81679 Muenchen, Germany
        }
\begin{document}
\date{}
\maketitle
\label{firstpage}
\pagerange{\pageref{firstpage}--\pageref{lastpage}} \pubyear{2015}
\begin{abstract}
We  perform multi-plane  ray-tracing  using  the GLAMER  gravitational
lensing  code within  high-resolution light-cones  extracted from  the
{\small CoDECS} simulations: a suite  of cosmological runs featuring a
coupling between Dark  Energy and Cold Dark Matter.  We  show that the
presence of the coupling is evident not only in the redshift evolution
of the  normalisation of the  convergence power spectrum, but  also in
differences  in   non-linear  structure  formation  with   respect  to
$\Lambda$CDM.  Using a tomographic approach  under the assumption of a
$\Lambda $CDM cosmology, we demonstrate that weak lensing measurements
would result  in a $\sigma  _{8}$ value  that changes with  the source
redshift if  the true  underlying cosmology is  a coupled  Dark Energy
one.  This provides a generic null test for these types of models.  We
also find that different models of coupled Dark Energy can show either
an enhanced or a suppressed  correlation between convergence maps with
differing source  redshifts as  compared to $\Lambda$CDM.   This would
provide  a  direct  way  to discriminate  between  different  possible
realisations of the coupled Dark Energy scenario.  Finally, we discuss
the  impact  of  the  coupling  on  several  lensing  observables  for
different source  redshifts and  angular scales with  realistic source
redshift distributions for current ground-based and future space-based
lensing surveys.
\end{abstract}

\section{Introduction}

It is now widely accepted by  the scientific community that the energy
content  of our  Universe must  be largely  dominated by  some unknown
particles and  fields beyond the  standard model of  particle physics.
These  are  characterised  by  extremely weak  interactions  with  the
electromagnetic field,  and are thereby termed  the ``dark" components
of the Universe \citep{suzuki11,planckxvi}, and generically classified
as ``dark  matter" and  ``dark energy" based  on their  background and
clustering properties.  While dark  energy is  supposed to  source the
observed            accelerated            cosmic            expansion
\citep{Riess_etal_1998,Schmidt_etal_1998,Perlmutter_etal_1999,kowalski08}
and to  have at  most weak spatial  density fluctuations,  dark matter
constitutes  more  than 80  percent  of  the  clustering mass  in  the
Universe    and   drives    the    growth    of   cosmic    structures
\citep{lacey93,tormen98a,springel05b,giocoli07a}   as   well  as   the
deflection  of   light  rays   as  predicted  by   General  Relativity
\citep{bartelmann01,bartelmann10}.  The latter  phenomenon is known as
``gravitational lensing": the light  traveling from background sources
(such as  distant galaxies) down to  the observer is deflected  by the
inhomogeneous intervening matter distribution, causing a distortion of
light bundles and  consequently a modification of  the observed galaxy
shape \citep{kaiser93,kaiser95}. Depending on the overall magnitude of
the light  deflection we can  distinguish between two main  regimes of
gravitational  lensing: the  strong  lensing  (SL) characterising  the
large distortions generated by single  highly overdense regions of the
Universe (as the central regions  of galaxies and galaxy clusters) and
the weak  lensing (WL) that occurs  as the integrated effect  of light
rays    traveling    through     the    inhomogeneous    cosmic    web
\citep{meneghetti05a,kneib11}.   In   particular,  weak  gravitational
lensing represents an important  and widely-used tool for cosmological
investigation  and to  probe the  matter density  around galaxies  and
galaxy  clusters \citep{mandelbaum06b,okabe10b,oguri12}.   However, as
this effect  is weak --  by definition --  it is necessary  to average
over  a  large  area  of  sky in  order  to  extract  a  statistically
significant lensing signal.

In the present work  we will focus on the prospects for  using WL as a
cosmological  probe  to distinguish  among  different  models of  dark
energy.  By analysing the distorted  shape of background galaxies as a
function of  their redshift, WL  can in principle constrain  the total
matter density of  the Universe, $\Omega _{\rm m}$,  the linear matter
power  spectrum  normalisation, $\sigma  _{8}$,  and  the dark  energy
equation  of  state  $w_{\rm  DE}$.  Furthermore,  this  could  reveal
possible signatures of an interaction  between the two dark components
\citep{beynon12,carbone13,pace15}.    In   order   to   achieve   such
discriminating  power, however,  highly  accurate  and unbiased  shear
measurements -- as  the ones expected for the next  generation of wide
surveys --  will be  required.  Since  gravitational lensing  does not
depend  on the  bias between  the distributions  of dark  and luminous
matter,  it represents  a  complementary probe  to other  constraining
observations like  supernovae, BAO (Baryon Acoustic  Oscillations) and
CMB       (Cosmic       Microwave       Background)       measurements
\citep{kilbinger13,kilbinger14}.

In recent  years, the  activities of  the CFHTLenS  collaboration have
greatly contributed to measuring the  cosmic shear signal in different
patches  of   the  sky   with  the   aim  to   constrain  cosmological
parameters. From the first exploration by \citet{fu08} of weak lensing
by large-scale structure in the linear regime on a region of 57 square
degrees,  the  CFHTLenS  work  has  progressed  through  a  series  of
different   steps   \citep{benjamin13,kilbinger13}  and   improvements
\citep{hildebrandt12,heymans12,heymans13}  before  performing  a  full
weak  lensing  measurement  in  three  dimensions  using  a  spherical
harmonic approach  \citep{kitching14}.  As  demonstrated by  the shape
measurement of  galaxies in  the COSMOS field  observed by  the Hubble
Space Telescope  \citep{amara12}, in  the near future,  WL space-based
measurements   \citep{refregier02,refregier04}  combined   with  other
independent probes of the large-scale matter distribution will be able
to discriminate with  high accuracy among various  possible scenarios for
the     fundamental     constituents     of    the     dark     sector
\citep{harnois-deraps14,kitching14b}.  Finally, WL is also a promising
tool  for  identifying  clusters  \citep{maturi05}  in  blank  fields,
complementing  and  possibly   driving  other  independent  approaches
\citep{bellagamba11}.

Ray-tracing through  light-cones extracted from  numerical simulations
represents  the  most  accurate   method  to  compute  WL  predictions
\citep{hilbert08}.  In  particular, this  is the case  for cosmologies
that have not yet been  fitted to semi-analytical prediction tools for
their behaviour in the non-linear regime \citep{camb,takahashi12}. The
drawback of this  approach is that it is  computationally demanding as
it requires  to sample  the full cosmological  parameter space  with a
series  of  N-body simulations.   However,  high  speed computers  and
improvements   of  numerical   solver  algorithms   are  progressively
simplifying the ray-tracing methodology.

Studying the  WL signal  signal as  a function  of source  redshift is
commonly called  ``lensing tomography''  and is  likely to  have great
significants for  cosmology in the future  \citep{schrabback10}.  This
is the technique we will investigate  in the present work with the aim
of understanding  whether a tomographic slicing  of background sources
might  increase  the  information  extracted from  WL  observables  on
possible  interactions  between  dark  matter  and  dark  energy.   In
particular, the possibility of fully exploiting this method to measure
specific dark energy signatures does not  only rely on a high accuracy
of the shear measurements, but also  on a high source density which is
necessary to reduce statistical errors.

The  {\small  CoDECS}   cosmological  simulations  are  hydrodynamical
simulations but  do not  include gas cooling,  star formation  and the
feedback mechanism of the baryonic  component, which may imprint a non
negligible signal in the cosmic shear power spectrum on angular scales
smaller           than            a           few           arcminutes
\citep{semboloni13,fedeli14,harnois-deraps14}.   However,  as  we  are
interested  in  examining relative  differences  of  the coupled  dark
matter-dark energy models  with respect to the  standard $\Lambda $CDM
one we will  make the common assumption that  the relative differences
between  models would  be only  weakly affected  by baryonic  physics.
This  assumption is  also reinforced  by  the fact  that the  baryonic
physics   has   a  non-negligible   impact   for   scale  $k   \gtrsim
3500\,h/\mathrm{Mpc}$.

\citet{pace15}  have  already  performed  ray-tracing  simulations  in
different coupled  dark matter-dark  energy models extracted  from the
{\small  CoDECS}  suite,  constructing   maps  for  different  lensing
quantities  starting from  the lensing  potential.  Their  analysis is
performed  on  maps  with  a  resolution of  around  $20$  arcsec  and
considering sources located at redshift  $z_s=1$.  They found that the
most significant  differences from the  standard $\mathrm{\Lambda}$CDM
model are due to differences in the growth of the perturbations and to
the effective friction term in  non-linear dynamics.  The most extreme
realisation  of  coupled Dark  Energy  expectedly  showed the  largest
difference  from $\mathrm{\Lambda}$CDM  of about  $40\%$ in  the power
spectrum, as found also by \citet{carbone13}.  In this paper we extend
and  complement their  analyses focusing  the attention  on the  small
scales  regime  thanks   to  high-resolution  ray-tracing  simulations
performed with the GLAMER code.   In addition we construct ray-tracing
simulations considering different  source redshift distributions, with
the  aim of  tomographically evaluate  the difference  between coupled
dark energy models and the standard $\mathrm{\Lambda}$CDM one.

\begin{figure}
\includegraphics[width=\hsize]{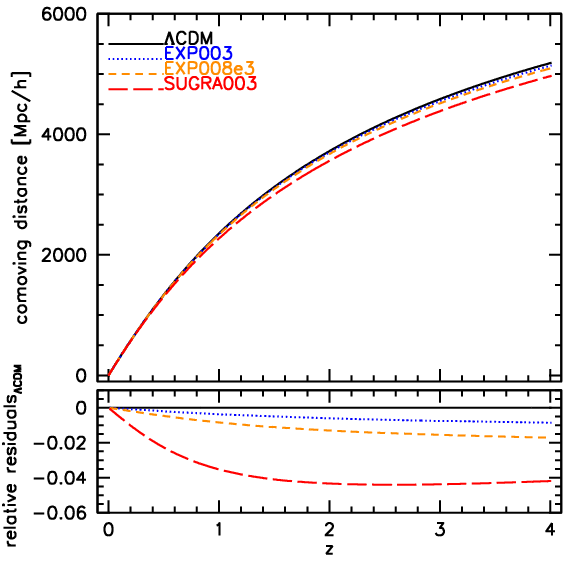}
\caption{Comoving distances  as a function  of redshifts for  the four
  considered models  of the  {\small CoDECS}  suite. The  bottom panel
  shows  the  relative  difference  of  the  models  with  respect  to
  $\Lambda\mathrm{CDM}$.\label{dldistance}}
\end{figure}

The  structure  of   this  paper  is  organized   as  follows.  In
Section~\ref{secnum} we  summarise the numerical simulations  that are
used to  construct the light-cones and  in Section~\ref{seclensing} we
describe   the  methodology   we   adopted   to  perform   multi-plane
ray-tracing. Results and lensing statistical analyses are presented in
Section~\ref{secresults}.   Finally,  in  Section~\ref{secsummary}  we
discuss our conclusions and summarise our main results.

\section{Cosmological Simulation}
\label{secnum}
\begin{table*}
\caption{The  list of  cosmological models  considered in  the present
  work and  their specific parameters.   All the models have  the same
  amplitude of scalar perturbations  at $z_{\rm CMB}\approx 1100$, but
  have different values of $\sigma _{8}$ at $z=0$.  In short, $\alpha$
  is a parameter in the  inflation potential as shown, $\beta(\phi)$ is
  the coefficient  of the coupling  term with dark matter  density and
  $w_{\phi  }(z=0)$  is  the  effective equation  of  state  parameter
  ($p/\rho$). See \citet{baldi12} for details.}
\label{tab:models}
\begin{tabular}{llcccc}
\hline
Model & Potential  &  
$\alpha $ &
$\beta (\phi )$ &
$w_{\phi }(z=0)$ &
$\sigma _{8}(z=0)$\\
\hline
$\Lambda $CDM & $V(\phi ) = A$ & -- & -- & $-1.0$ & $0.809$ \\
EXP003 & $V(\phi ) = Ae^{-\alpha \phi }$  & 0.08 & 0.15 & $-0.992$ & $0.967$\\
EXP008e3 & $V(\phi ) = Ae^{-\alpha \phi }$  & 0.08 & $0.4 \exp [3\phi ]$& $-0.982$ & $0.895$ \\
SUGRA003 & $V(\phi ) = A\phi ^{-\alpha }e^{\phi ^{2}/2}$  & 2.15 & -0.15 & $-0.901$ & $0.806$ \\
\hline
\end{tabular}
\end{table*}

\begin{figure*}
\includegraphics[width=\hsize]{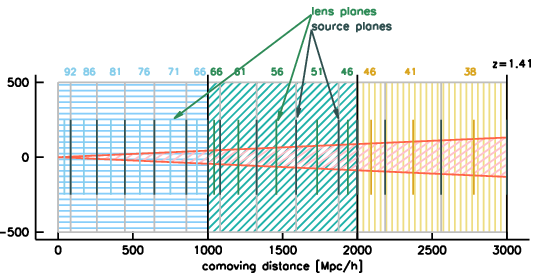}
\caption{An illustration of  the construction of the  light-cone up to
  redshift $z=1.4$ with an aperture  of $5\times 5$ square degrees for
  the  $\mathrm{\Lambda}$CDM  model.   The light-cone  passes  through
  three simulation boxes, since the comoving distance corresponding to
  $z=1.4$  is $3$  Gpc/$h$.  As  indicated in  the upper  part of  the
  figure we have $12$ snaphosts available up to this considered source
  redshift from  which we construct  $14$ lens planes.  In  the square
  that contains a single simulation  box we stack together slices from
  difference snapshots of the box.   The snapshot id numbers are shown
  on the top of each rectangle representing the slices.  The different
  colors used to shade the  squares indicate that the simulation boxes
  have been randomized as described in \citet{roncarelli07}.  The pink
  shaded triangle shows the region of the simulation snapshots used to
  construct          the           planes          within          the
  light-cones.\label{lightconedescription}}
\end{figure*}

For  our  investigation we  rely  on  the publicly  available  {\small
  CoDECS} simulations  \citep[][]{baldi12} that represent  the largest
suite of  cosmological and hydrodynamical simulations  of coupled Dark
Energy (cDE) models  to date.  The simulations have  been performed by
means  of a  modified  version developed  by  \citet{baldi10}, of  the
widely    used     TreePM/SPH    N-body    code     {\small    GADGET}
\citep[][]{springel05a}, and self-consistently include all the effects
associated with the non-minimal interaction  between a DE scalar field
$\phi $ and CDM particles.  The {\small CoDECS} suite includes several
different  possible combinations  of the  scalar DE  potential --  the
exponential    \citep[][]{lucchin85,wetterich88}    or    the    SUGRA
\citep[][]{brax99}  potentials  for example  --  and  of the  coupling
function which  can be  either constant or  exponential in  the scalar
field  \citep[see  e.g.][]{baldi11}.  In  the  present  work, we  will
consider four  models ($\Lambda $CDM, EXP003,  EXP008e3, and SUGRA003)
with  different   combinations  of  these  free   functions  that  are
summarised in Table~\ref{tab:models}.  For  more details on the models
we refer to \citet{baldi12}.

Such variety  of scenarios  is reflected in  the diversity  of effects
that they determine  on both the background expansion  history and the
linear and non-linear evolution  of perturbations.  More specifically,
the  dynamical evolution  of  the  DE scalar  field  through a  matter
dominated scaling  solution alters  the cosmic expansion  history with
respect  to  the  standard  $\Lambda  $CDM cosmology  in  a  way  that
substantially depends on the potential and coupling functions adopted.
This  is shown  in Fig.~\ref{dldistance},  where we  show the  comoving
distance as  a function of redshift  for the four models.   The bottom
panel displays the relative differences of the comoving distances at a
given redshift between the cDE  models and the reference $\Lambda$CDM.
From  the  figure  we  can  notice that  while  locally  the  comoving
distances  are   consistent  with   each  other   (due  to   the  same
normalisation of Hubble parameter,  $H_{0}$, for all the cosmologies),
at higher redshifts the cDE  models have smaller volumes than $\Lambda
$CDM, with a maximum effect  for the SUGRA003 cosmology which deviates
low from $\Lambda $CDM by about 5 percent already at $z=2$.

At the level  of linear density perturbations, the  models all predict
an  enhanced  growth  rate  with  respect to  $\Lambda  $CDM  at  high
redshifts. However, while  the EXP003 and the EXP008e3  models show an
enhanced growth also  at low redshifts, thereby resulting  in a larger
value  of   $\sigma  _{8}$  at   $z=0$,  the  SUGRA003   cosmology  is
characterized  by a  slower growth  as compared  to $\Lambda  $CDM for
$z\lesssim 7$  resulting in a  comparable value of $\sigma  _{8}$ (see
the last column of Table~\ref{tab:models}).  The non-linear effects of
these models  have been studied  in several publications based  on the
outcomes of the {\small CoDECS}  simulations and range from the impact
of  cDE   on  the  abundance   and  structural  properties   of  halos
\citep[][]{baldi12a,baldi11a,cui12,giocoli13},   on  the   statistical
properties    of     the    large-scale     structures    distribution
\citep[][]{marulli12,moresco14},    on   the    properties   of    the
Inter-Galactic Medium  at high redshifts \citep[][]{baldi10b},  and on
weak lensing statistics  \citep[][]{beynon12,carbone13,pace15}. In the
present work, we aim at extending the latter analysis by investigating
whether a  tomographic slicing  of the  background sources  within the
light-cone  of  a  25  square  degrees field  of  view  might  provide
additional information  to observationally distinguish the  cDE models
from the standard  $\Lambda $CDM cosmology and  possibly the different
cDE models from each other.

For our  analysis we  will make  use of  the {\small  L-CoDECS} series
consisting of a periodic cosmological  box of $1$ Gpc$/h$ aside filled
with  $2\times  1024^{3}$   particles  evolved  through  collisionless
dynamics from $z=99$  to $z=0$. All the models share  the same initial
conditions at the  redshift of the CMB $z_{\rm CMB}  \approx 1100$ and
have the  same cosmological  parameters at  $z=0$ consistent  with the
WMAP7      cosmological     results      \citep{komatsu11},     namely
$\Omega_{\mathrm{CDM}}=0.226$,           $\Omega_{\mathrm{DE}}=0.729$,
$h=0.703$,   $A_s=2.42   \times    10^{-9}$,   $\Omega_b=0.0451$   and
$n_s=0.966$.  The  mass resolution  is $M_{\rm CDM}(z=0)  = 5.84\times
10^{10}$ M$_{\odot }/h$ for CDM  particles and $M_{\rm b} = 1.17\times
10^{10}$  M$_{\odot }/h$  for the  (collisionless) baryonic  particles
\citep[see][for   a    detailed   discussion]{baldi12},    while   the
gravitational softening was set to $\epsilon _{g} = 20$ kpc$/h$.

\section{Lensing Pipeline}
\label{seclensing}

In the following sections we will present the procedure we followed in
constructing  the  lens  planes  from  the  cosmological  simulations,
assembling  them  into light-cones  and  tracing  the paths  of  light
through them.

\subsection{Constructing the light-cone: MapSim}

Our code for  extracting the particles from  the simulation's snapshot
files and  assembling them into  a light-cones is called  MapSim.  The
steps  MapSim  goes  through  in  constructing  a  light-cone  can  be
summarized as follows:
\begin{itemize}
\item Read in an input  parameter file that contains information about
  the desired  field of  view, highest source  redshift (in  this case
  taken to be $z_s=4$) and locations of snapshot files.  The number of
  lens planes required is decided ahead of time in order to avoid gaps
  in the constructed light-cone.  The  choice $z_s=4$ has been made to
  better understand  where the  dynamical evolution  of the  DE scalar
  field and  the enhanced growth  rate --  in the different  models --
  start to leave  a mark in the weak lensing  observables. Notice that
  at low redshifts the models show a different behaviour: while EXP003
  and EXP008e3 continue their enhanced growth, SUGRA003 does not.
\item Read in each snapshot file going from the present time to higher
  redshift  snapshots while  extracting  only  the particle  positions
  within the  desired field  of view.   Only a  single snapshot  is in
  memory at any time.
\item  Selection and  randomization of  each  snapshot is  done as  in
  \citet{roncarelli07}.   If the  light-cone reaches  the border  of a
  simulation box before it has reached a redshift range where the next
  snapshot will be  used, the box is re-randomized  and the light-cone
  extended through it again.
\item The lensing  planes are built by mapping  the particle positions
  to the nearest pre-determined  plane, maintaining angular positions,
  and then pixelizing the surface  density using the triangular shaped
  cloud (TSC) method \citep{hockney88}.  The grid pixels are chosen to
  have the same angular size on  all planes. The lens planes have been
  constructed each time a piece of simulation is taken from the stored
  particle snapshots; their number and  frequency depend on the number
  of snaphosts stored while running the simulation.
\end{itemize} 
In  Fig.~\ref{lightconedescription} we  show  an  illustration of  the
construction  of  the  light-cone  up to  redshift  $z=1.4$,  for  the
$\mathrm{\Lambda}$CDM  model, piling  one  on top  of  the others  the
different  portions  of  the simulation  snapshots.   The  differently
colored   squares   are   different  realizations   using   the   same
randomization method.   The rectangles  within the  squares represents
the portion of the simulation snapshot from which the particle density
distributions have been  taken. The number on the top  of them are the
id numbers  of the  corresponding simulation snapshots.   The vertical
line in the  middle of each rectangle indicates the  planes onto which
the particles  are projected for  ray-shooting.  We remind  the reader
that to  save disk space not  all $92$ simulation snapshots  have been
stored in  running the  simulation.  Up to  redshift $z_s=4$  we saved
$18$ snapshots which  are sufficient to consistently  model the matter
density distribution in our light cones.

The maps were  constructed with a $5\times 5$ sq.   deg. field of view
and an  angular resolution of  $8.8$ arcsec.   For each model  we have
constructed twenty-five  independent realizations, being  careful that
in  each  realization the  same  field  of  view  is selected  in  the
different  cosmologies.  However,  considering the  different comoving
distance-redshift evolution  through the simulation snapshots  we have
built 22  planes up to  redshift $z=4$ for  the $\mathrm{\Lambda}$CDM,
EXP003 and EXP008e3,  while we have 21 planes for  the SUGRA003 model.
In   Table~\ref{tabdist}  we   point   out   the  comoving   distances
corresponding to redshift $z_s=0.5$, 1.4 and 4 for the four models.

\begin{table}
\caption{Comoving  distances in  Mpc$/h$  for  three different  source
  redshifts in the considered cosmological models}
\label{tabdist}
\begin{tabular}{l|c|c|c|}
\hline
model & $z_s=0.5$ & $z_s=1.4$ & $z_s=4$   \\ \hline
$\mathrm{\Lambda}$CDM & 1327.27 & 3000 & 5179.52   \\
EXP003 & 1324.56 & 2999.75 & 5135.05   \\
EXP008e3 & 1320.86 & 2981.97 & 5091.15   \\
SUGRA003 & 1297.07 & 2892.85 & 4962.47   \\
\hline
\end{tabular}
\end{table}

\begin{figure*}
\includegraphics[width=0.75\hsize]{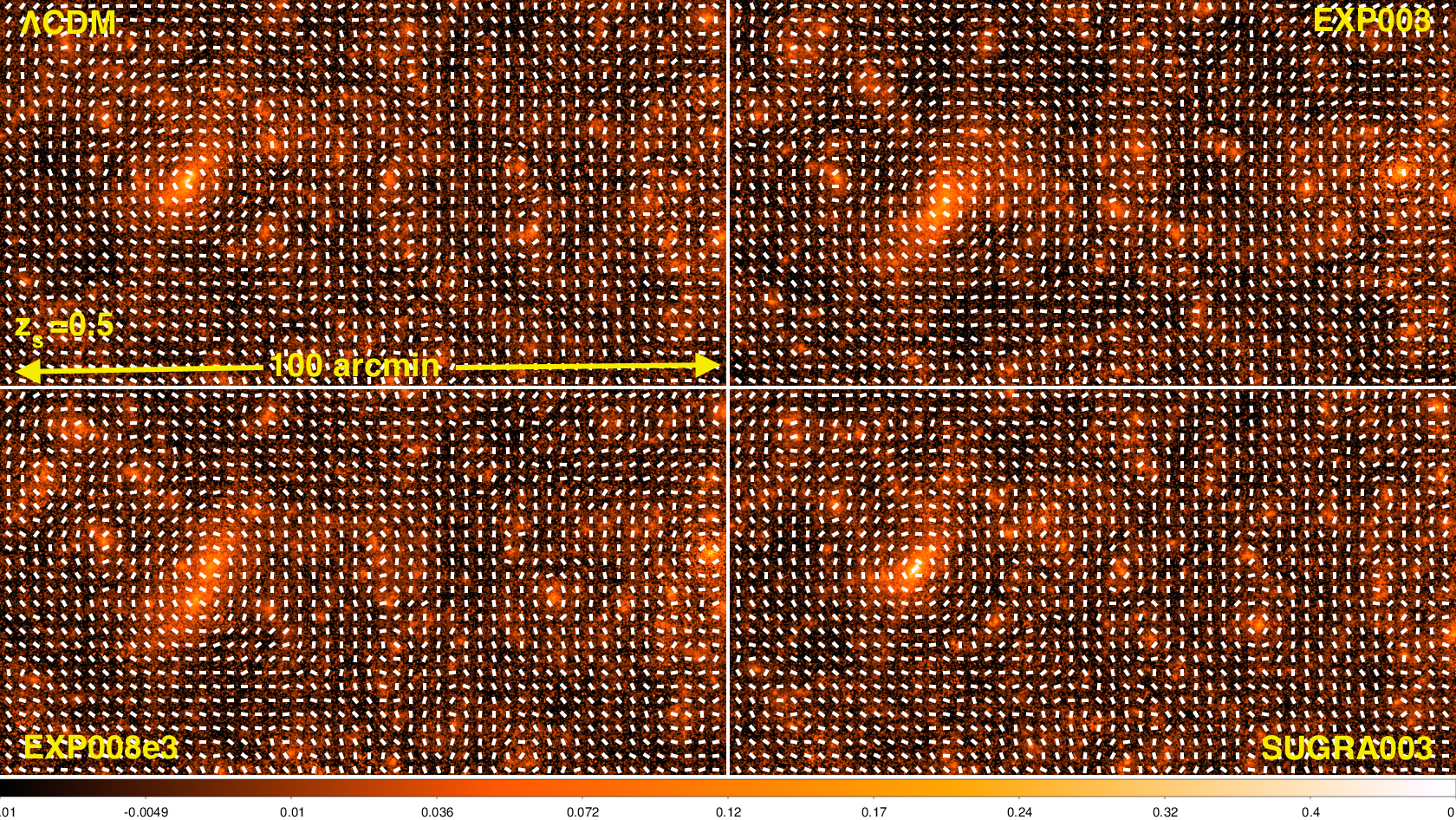}
\includegraphics[width=0.75\hsize]{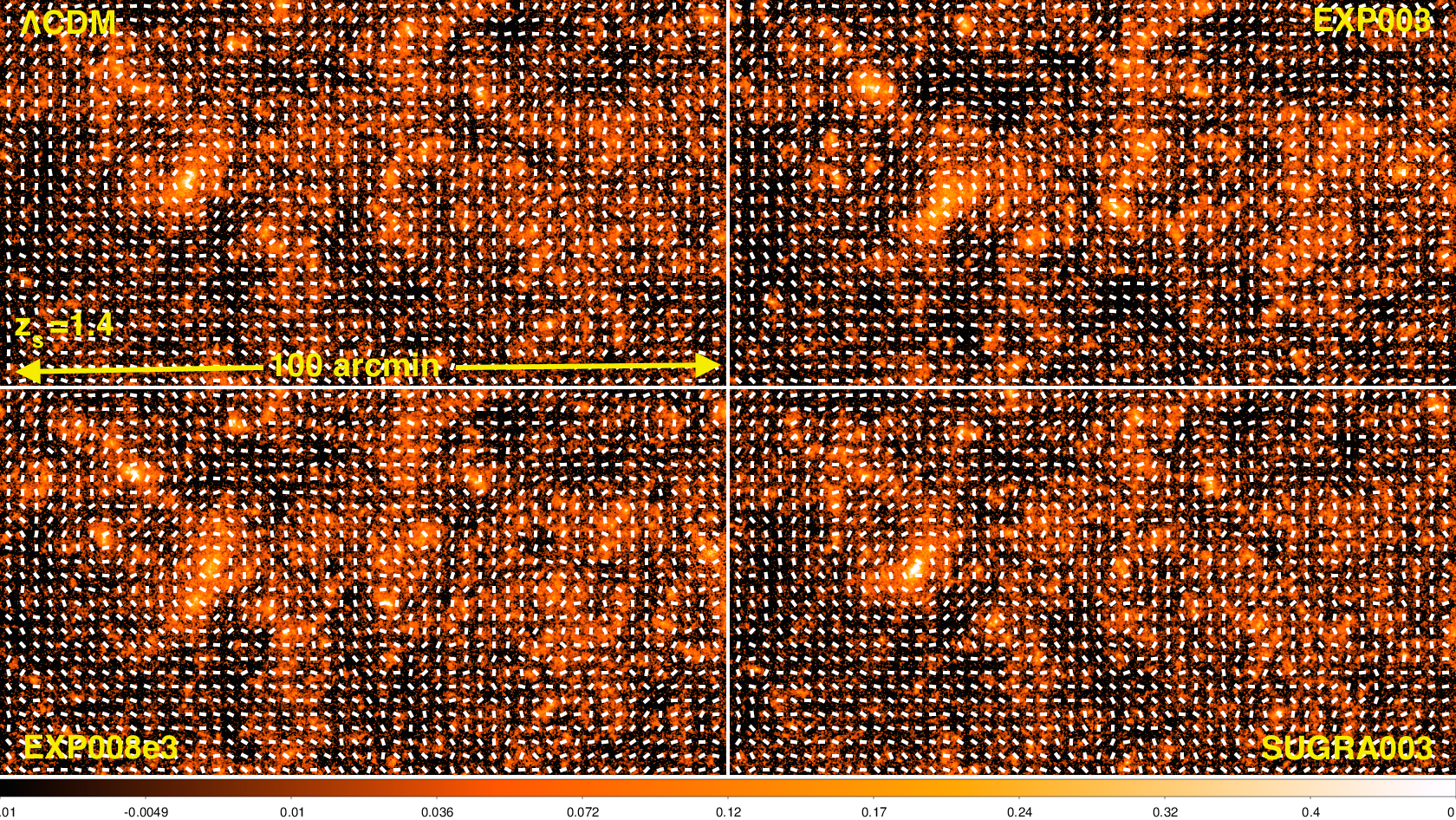}
\includegraphics[width=0.75\hsize]{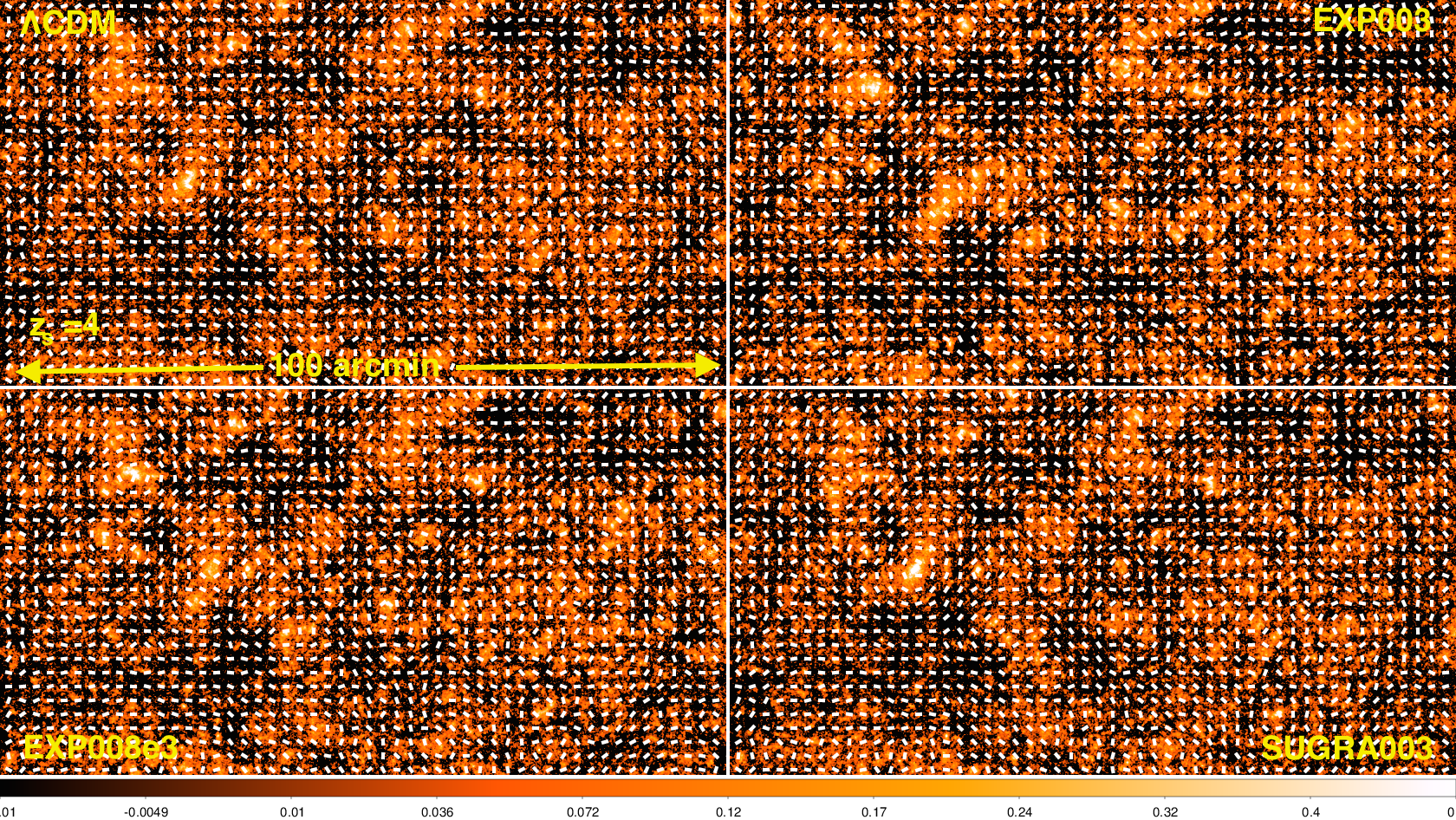}
\caption{Convergence maps of a light-cone  with aperture 100 arcmin on
  the x-side, for the four considered models -- in each panel top left
  $\mathrm{\Lambda}$CDM, top  right EXP003,  bottom left  EXP008e3 and
  bottom right  SUGRA003. The three  panels refer to  different cases:
  sources located  at redshifts $z_s=0.5,\,1.4$  and $4$, from  top to
  bottom, respectively. In  each map the sticks show  the direction of
  the corresponding field.
    \label{figkappamaps}}
\end{figure*}

\subsection{Ray-Tracing through the planes}

Once the lens planes are created  as described in the previous section
the lensing calculation  itself is done using the  GLAMER lensing code
\citep{metcalf13,petkova13}.   The  multiplane ray-tracing  method  is
described in  detail in \cite{petkova13}  so we will only  outline the
procedure here.

A few definitions are required.  If the angular position on the sky is
${\pmb \theta}$ and  the position on the source plane  expressed as an
angle (the  unlensed position)  is ${\pmb  \beta}$, then  a distortion
matrix ${\bf A}$ can be defined as
\begin{align}
{\bf A} \equiv \frac{\partial {\pmb \beta}}{\partial {\pmb \theta} } = 
\left( 
\begin{array}{cc}
1-\kappa-\gamma_1 & \gamma_2 - \gamma_3 \\
\gamma_2 + \gamma_3& 1-\kappa + \gamma_1
\end{array}
\right)\ .
\end{align}
The traditional decomposition of this  matrix is shown, where $\kappa$
is called  the convergence and  ${\pmb \gamma}$ represents  the shear.
The torsion,  $\gamma_3$, represents a  rotation which can  occur when
there  are multiple  deflection  planes.   It is  of  order the  shear
squared  \citep[see][]{petkova13}  and   according  to  our  numerical
calculations, and those of \cite{becker13},  it is quite small, but it
will be retained here for completeness.

\begin{figure}
\includegraphics[width=\hsize]{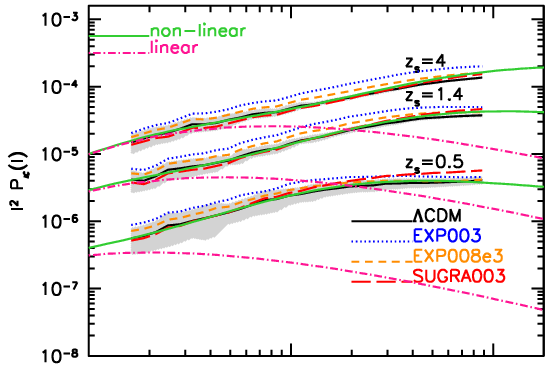}
\includegraphics[width=\hsize]{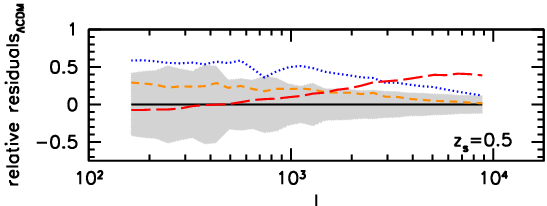}
\includegraphics[width=\hsize]{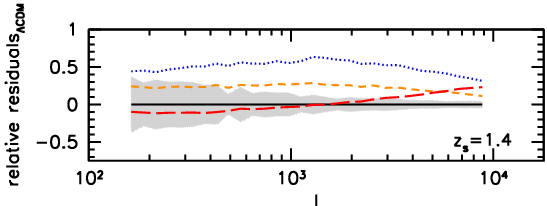}
\includegraphics[width=\hsize]{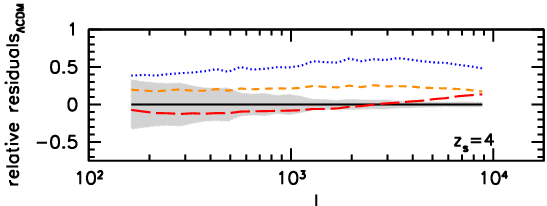}
\caption{Convergence  power spectrum  of the  four considered  {\small
    CoDECS}   cosmologies  for   three  different   source  redshifts:
  $z_s=0.5,\,1.4$  and  $4$.   The  shaded grey  region  encloses  the
  standard  deviation   of  the   mean  associated  to   the  $\Lambda
  \mathrm{CDM}$ model on the different  $5 \times 5$ degree light-cone
  realisations.  In the top panel the solid and dot-dashed curves show
  the  linear  and  non-linear   CAMB  predictions  for  the  $\Lambda
  \mathrm{CDM}$  cosmology  respectively.   For the  non-linear  power
  spectrum prediction  we adopt  the extended  version of  the Halofit
  Model  \citep{smith03}  from  \citet{takahashi12}.   In  the  bottom
  panels we  present the relative  residuals of the  convergence power
  spectra with respect to the  $\Lambda \mathrm{CDM}$ cosmology of the
  three  cosmological models  --  for each  of  the considered  source
  redshift cases  -- featuring a  direct interaction between  the Dark
  Energy and Dark Matter.\label{figpowerl}}
\end{figure}

When there is a single lens plane, the convergence can be expressed as a
dimensionless surface density,
\begin{equation}\label{eq:magnification_matrix}
\kappa({\pmb \theta}) \equiv \frac{\Sigma({\pmb \theta})}{\Sigma_{\rm
    crit}}\ ,
\end{equation}
where 
\begin{equation}
\Sigma_{\rm crit} \equiv \dfrac{c^2}{4 \pi G} \dfrac{D_l}{D_s D_{ls}}
\end{equation}
is called  the critical  density, $c$  is the speed  of light,  $G$ is
Newton's  constant  and  $D_l$  $D_s$ and  $D_{ls}$  are  the  angular
diameter   distances   between  observer-lens,   observer-source   and
source-lens,  respectively.  In  general, with  multiple lens  planes,
this is not the case however.

The deflection caused by a lens  plane, ${\pmb \alpha}$, is related to
the  surface density  on  the plane,  $\Sigma({\bf  x})$, through  the
differential equations
\begin{align}\label{eq:single_plane_kappa}
\nabla^2 \phi({\bf x}) = \frac{4 \pi G}{c^2} \Sigma({\bf x})   ~~~,~~~
  {\pmb \alpha}({\bf x}) = \nabla  \phi({\bf x}) .
\end{align}
where the  derivatives are with  respect to  the position on  the lens
plane.  These equations are solved  on each source plane by performing
a Discrete Fourier Transform (DFT)  on the density map, multiplying by
the appropriate factors and then transforming back to get a deflection
map with  the same resolution as  the density map.  With  the same DFT
method the  shear caused by  each plane is  simultaneously calculated.
Since  the  rays are  propagated  between  planes using  the  standard
distances  in  a  Robertson-Walker  metric  which  assumes  a  uniform
distribution of  matter the addition of  matter on each of  the planes
will,  in a  sense,  over-count  the mass  in  the universe.   Without
correcting for  this the average  convergence from the planes  will be
positive and will  cause the average distance for a  fixed redshift to
be smaller than it should be.   To compensate for the implicit density
between  the planes  the ensemble  average  density on  each plane  is
subtracted.  Each plane  then has zero convergence on  average and the
average redshift-distance  relation is as  it would be in  a perfectly
homogeneous universe.

After the deflection  and shear maps on each plane  are calculated the
light-rays are traced  from the observers through the  lens planes out
to the  desired source redshift.   The shear and convergence  are also
propagated through the planes as detailed in \cite{petkova13}.  GLAMER
performs a  complete ray-tracing  calculation that takes  into account
non-linear coupling terms  between the planes as  well as correlations
between the deflection  and the shear.  No weak  lensing assumption is
made at this stage.  The rays are shot in a grid pattern with the same
resolution as the mass maps: 5  degrees resolved with 2048 pixels on a
side.

In Fig.~\ref{figkappamaps}  we show the  convergence maps of  the same
light-cone realisation extracted from the different models -- top left
$\mathrm{\Lambda}$CDM,  top right  EXP003,  bottom  left EXP008e3  and
bottom right SUGRA003.  We show the  maps for sources located at three
fixed  redshifts:   $z_s=0.5$,  1.4   and  4   from  top   to  bottom,
respectively.  The sticks in each panel indicate the directions of the
corresponding  shear   field.   As   discussed  in   \cite{cui12}  and
\citet{giocoli13}, we immediately notice that the density distribution
differs in cDE  models from $\mathrm{\Lambda}$CDM due  to a difference
in the  growth as  a function  of redshifts. The  top panels  show the
presence  in the  field of  view of  a cluster,  at redshift  $z<0.5$,
which, while  it appears ``assembled'' and  with a single peak  in the
$\mathrm{\Lambda}$CDM  and the  SUGRA003 models,  it is  less evolved,
showing multiple components, in EXP003 and EXP008e3.  The intermediate
and the high redshift maps  also exhibit differences mainly because of
differences in the  evolution of the power  spectrum normalisation and
non-linear structure formation.

\subsection{Analytic methods}

An approximation is  commonly used to calculate  the convergence power
spectrum that avoids  the complications discussed in  the previous two
sections.
The  convergence can  be  calculated  by adding  up  the single  plane
convergences,    equation~\eqref{eq:single_plane_kappa},   along    an
unperturbed light-ray,  the Born approximation.  This  results in the
expression
\begin{equation}
\kappa(z_s,\pmb{\theta}) = \dfrac{3  H^2_0 \Omega_m}{2 c^2} \int_0^{w(z_s)} \mathrm{d}w
\dfrac{D(z)D(z,z_s)}{D(z_s) a(z)^2} ~ \delta\left(D(z)\,
  \pmb{\theta},z\right)\ ,
\end{equation}
where $w$  is the radial  comoving distance, $a \equiv  (1+z)^{-1}$ is
the  scale  factor  and  $\delta({\bf  x})  \equiv  (\rho({\bf  x})  -
\overline{\rho})/\overline{\rho}$  is   the  density   contrast.   The
angular power  spectrum of $\kappa$ in  the small angle limit  is then
found to be
\begin{equation}
P_{\kappa}(l) = \dfrac{9 H_0^4 \Omega_m^2}{4 c^4} \int_0^{w_s} 
\mathrm{d}w \left( \dfrac{D(z,z_s)}{D(z_s) a(z)}\right)^2 P_{\delta}
\left( l \dfrac{a(z)}{D(z)},z\right)
\label{eqpowerkappa}
\end{equation}
\citep{Kaiser_1992}.  Either  an analytic model for  the density power
spectrum, $P_{\delta}(k,z)$, or a power spectrum taken directly from a
simulation can be inserted into eq.\eqref{eqpowerkappa}.

In the  following sections  we will  contrast our  direct calculations
with  some  analytic  models   using  this  approximation.   For  some
applications this approximation is adequate, but in new situations and
for particular statistics it needs to be checked against simulations.

\section{Results}
\label{secresults}
\subsection{Convergence Power Spectrum}

\begin{figure*}
\includegraphics[width=0.45\hsize]{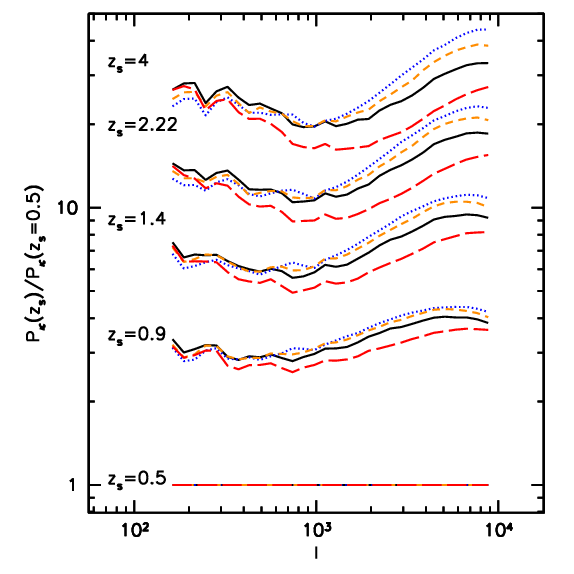}
\includegraphics[width=0.45\hsize]{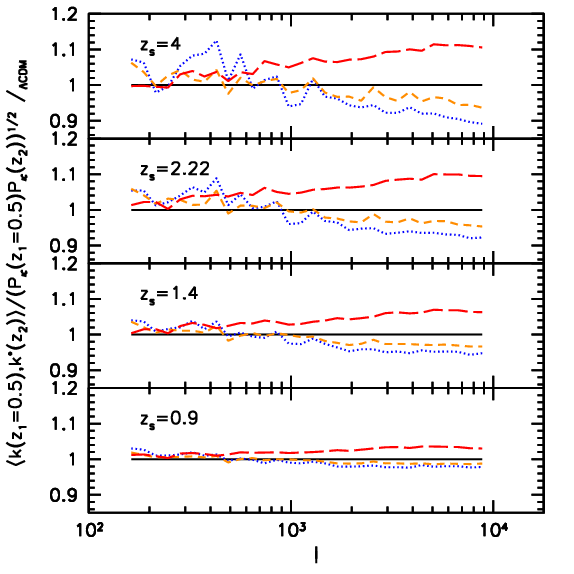}
\caption{Left  panel: ratio  between  the  convergence power  spectrum
  computed at different source redshifts -- as indicated in the labels
  -- and  the one  at $z_s=0.5$  for each  cosmological model.   Right
  panel:  rescaled cross  power spectra  between redshift  $z=0.5$ and
  $z_2$ --  as indicated in  the labels  -- for the  different models,
  computed from  the lensing  maps.  All the  cross spectra  have been
  rescaled with respect to the $\mathrm{\Lambda}$CDM prediction.  This
  is why  the correlation parameter on  the right can be  greater than
  one.    Different   line    styles    and   colors    are   as    in
  Fig.~\ref{figpowerl}. \label{figpowerlratio}}
\end{figure*}

In the real  (angular) space, a direct measurement of  weak lensing is
the two-point shear correlation functions $\xi_+$ and $\xi_-$ that can
be obtained  by averaging over  galaxy pairs with  angular separations
$|\pmb{\theta}_i - \pmb{\theta}_j|$ within a bin $\theta$:
\begin{equation}
  \xi_{\pm}(\theta) = \dfrac{\sum_{ij} w_i w_j \left[ \epsilon_t(\pmb{\theta}_i)
      \epsilon_t(\pmb{\theta}_j) \pm \epsilon_{\times}(\pmb{\theta}_i)
      \epsilon_{\times}(\pmb{\theta}_j) \right]}{\sum_{ij}w_i w_j}\ ,
\end{equation}
where the  measured galaxy  ellipticity measurements  $\epsilon_t$ and
$\epsilon_{\times}$  are  the  tangential and  cross  components  with
respect to  the line connecting  the pair, respectively.   The weights
$w$  are   obtained  from   the  galaxy  shape   measurement  pipeline
\citep{berstein02,bridle10,kacprzak12,miller13}.     These   two-point
shear  correlation functions  can be  calculated from  the convergence
power spectrum by the relation:
\begin{equation}
\xi_{+/-} = \dfrac{1}{2 \pi} \int_0^{\infty} \mathrm{d} l~ l
P_{\kappa}(l) J_{0/4}(l\theta)\,,
\end{equation}
where $J_0$  and $J_4$ are  the Bessel functions  and we have  set the
B-mode power spectrum  to zero because lensing  generates only E-modes
in the weak lensing limit.

\begin{figure*}
\includegraphics[width=0.49\hsize]{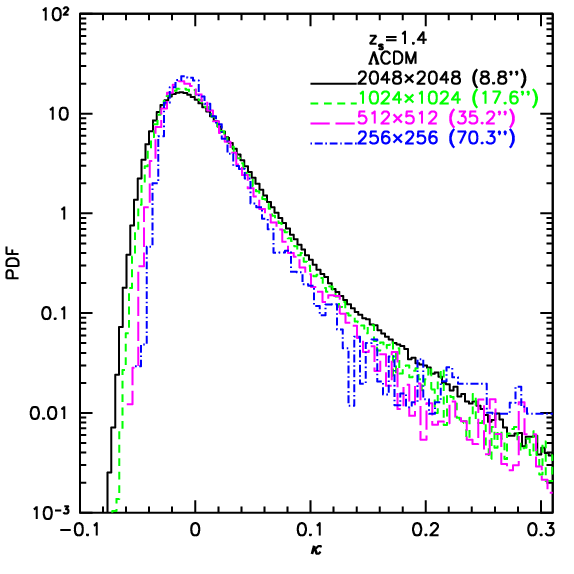}
\includegraphics[width=0.49\hsize]{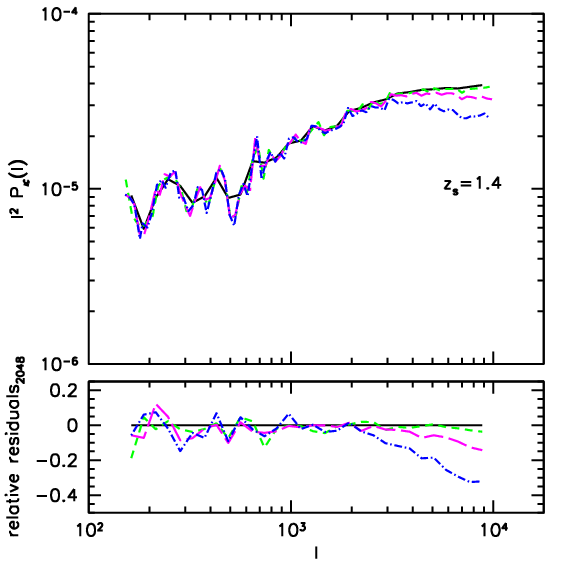}
\caption{Left  panel:  histograms of  the  convergence  values in  one
  realization of the  $\mathrm{\Lambda}$CDM model, considering sources
  at $z_s=1.4$.  The solid black histogram  shows the case for  the map
  resolution  used throughout  this paper  ($8.8$ arcsec  pixels). The
  short  dashed  green,  long   dashed  magenta  and  dot-dashed  blue
  histograms show the PDF of the maps degraded by a factor of 2, 4 and
  8, respectively.  Right top panel:  convergence power spectra of 
  these maps. Right bottom panel: the ratios of the power spectra with
  different  resolutions  with  respect   to  the  highest  resolution
  one.\label{figtestPDF}}
\end{figure*}

In Fig.~\ref{figpowerl} we  show the convergence power  spectrum up to
$l\approx 10^4$ measured in  the four cosmologies, considering sources
at  three  different  redshifts:  $z_s=0.5$, 1.4  and  4.   The  curve
referring  to  each   model  represents  the  average   over  all  the
light-cones.   For the  $\mathrm{\Lambda}$CDM case  the shaded  region
encloses  the  standard  deviation  of  the  mean  associated  to  the
different realizations.   The $l_{min}$ considered corresponds  to the
minimum resolvable in the assumed field of view. In the same figure we
also  show  the  predictions   obtained  by  inserting  into  equation
(\ref{eqpowerkappa}) analytic models for the linear and the non-linear
power  spectrum,   namely  the  Halofit  Model   \citep{smith03}  from
\citet{takahashi12}  implemented in  CAMB \citep{camb}.   As shown  in
\citet{pace15} -- where they  consider the matter density distribution
only up  to redshift $z_s=1$ --  the EXP003 model has  much more power
than the  $\mathrm{\Lambda}$CDM with  the largest difference  being on
small scales and increasing  with the source redshift.  Interestingly,
for   the    SUGRA003   model,    the   power   spectrum    is   below
$\mathrm{\Lambda}$CDM  for  low  multipoles  and  above  it  for  high
multipoles.  This  is a  result of  the fifth force  term in  the dark
matter-dark energy coupling which  drives rapid structure formation at
high redshift, but slows it down  at lower redshift.  The SUGRA003 and
the  $\mathrm{\Lambda}$CDM power  spectra  intersect  at about  $l\sim
10^3$,   slightly  decreasing   with  the   source  redshifts.    From
Fig.~\ref{figpowerl} we  also see  that the  EXP008e3 model  has about
15-25 percent more power than  $\mathrm{\Lambda}$CDM at all scales and
source redshifts, a signature of its growth rate being maintained over
the whole  redshift range.   As obtained  by the  tomographic analysis
performed by \citet{pace15}, where only the ratio of the power spectra
for sources  at redshift $z_s=2$  and $z_s=1$ has been  considered, we
confirm that the  small differences they find at very  high $l$ in the
coupled models, and  particularly in the bouncing  model SUGRA003, are
actually  present   and  are  much  clearer   in  our  high-resolution
ray-traced maps.  The same behaviour of  the power spectra we find has
also been noticed and discussed by \citet{carbone13} using ray-tracing
technique   to   compute   the   CMB  lensing   maps;   however   also
\citet{carbone13} are not able resolve the small scale features of the
coupled DE-DM models  because the angular resolution on  their maps is
more than a factor of ten lower than ours.

Clearly weak lensing  tomography, i.e.  the study of  the weak lensing
signal as a function of the  source redshift, can be an important tool
for  studying these  differences in  the evolution  of structure  with
redshift  that   occur  in  different  coupled   Dark  Energy  models.
Fig.~\ref{figpowerlratio} shows this more  clearly.  In the left panel
we display the  ratio between the convergence  power spectrum computed
at five  different source redshifts  and the one computed  for sources
located  at $z_s=0.5$;  line  styles and  colors are  the  same as  in
Fig.~\ref{figpowerl}.  For EXP003 and EXP008e3  the ratios tend to lie
above the $\mathrm{\Lambda}$CDM one, but  for SUGRA003 it stays below.
This is  a result of  structures in  the SUGRA003 model  evolving less
rapidly at  late times while  in the  EXP003 and EXP008e3  models they
evolve  more  rapidly.  The  difference  between  the models  is  most
evident  at small  scales  as a  consequence  of non-linear  structure
formation.   The right  panel  of  Fig.~\ref{figpowerlratio} show  the
rescaled cross-spectra between the  convergence at different redshifts
within  the  same light-cone.   High  correlation  indicates that  the
lensing is being caused by  the same objects.  Interestingly, SUGRA003
has more correlations at small scales  and less at large scales, while
the other  models have the  opposite trend,  due to an  enhancement of
their  growth  rate at  high  redshifts  and  to  a depletion  at  low
redshifts with respect to the $\mathrm{\Lambda}$CDM one. We remind the
reader that the growth rate in  EXP003 and EXP008e3 is always enhanced
with respect to the standard model  causing a higher $\sigma_8$ at the
present time.

\begin{figure*}
\includegraphics[width=0.45\hsize]{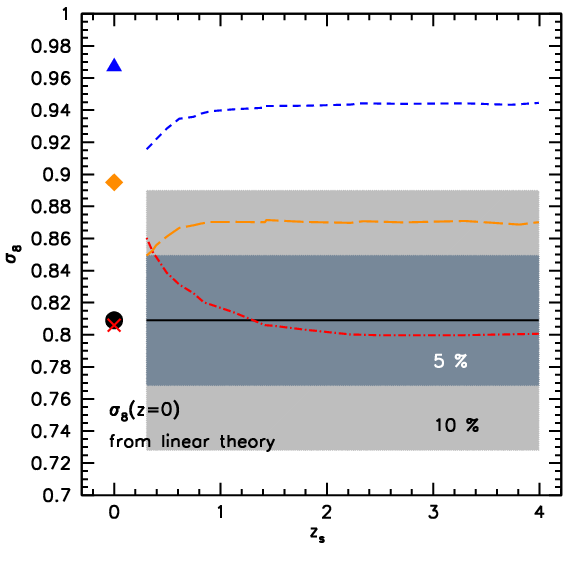}
\includegraphics[width=0.45\hsize]{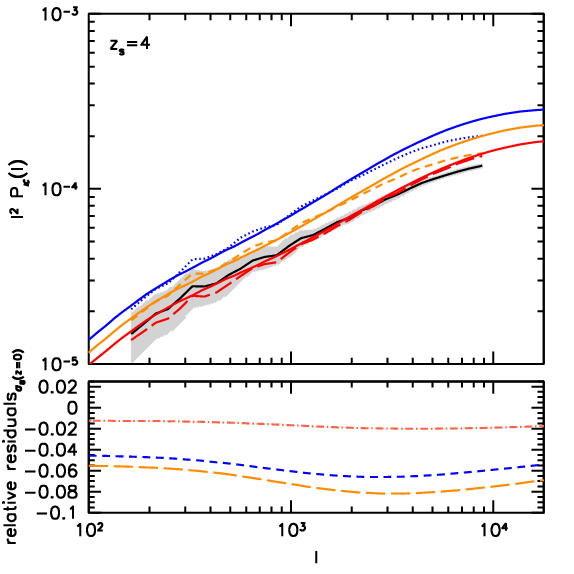}
\caption{Left panel: $\sigma_8$ as a  function of the source redshift,
  obtained by fitting the corresponding convergence power spectrum for
  each  cosmological model,  fixing the  total mass  density parameter
  $\Omega_m$ to its true value  in the simulation.  The colored points
  show the value  of $\sigma_8$ at the present  time interpolated from
  linear theory:  blue triangle (EXP003) ,  orange diamond (EXP008e3),
  red cross  (SUGRA003) and  black dot  ($\mathrm{\Lambda}$CDM). Right
  top panel: convergence power spectra for sources at redshift $z_s=4$
  as measured  from the light-cones  in the four  cosmological models.
  As in  previous figures solid,  dotted, dashed and  dot-dashed lines
  correspond to $\mathrm{\Lambda}$CDM,  EXP003, EXP008e3 and SUGRA003,
  respectively.  The corresponding solid  colored curves represent the
  best $\mathrm{\Lambda}$CDM  power spectra obtained fixing  the total
  matter  content and  varying  $\sigma_8$.  Bottom  right panel:  the
  ratio  between   the  convergence  power  spectrum   with  the  best
  $\sigma_8$  and the  one obtained  using $\sigma_8(z=0)$,  i.e.  the
  interpolated value using linear  theory \citep{baldi12} is shown for
  the three cDE models.
\label{figBP}}
\end{figure*}

The pixel resolution of our maps ($8.8$ arcsec in all cases) will have
some impact on the accuracy  of the lensing statistics calculated.  In
particular the  pixelation tends  to smooth out  peaks and  reduce the
power on small scales  \citep{takahashi11,pace15}.  To investigate how
our pixelation might be affecting the results we constructed the pixel
Probability  Distribution Function  (PDF)  of the  convergence map  as
shown   in   Fig.~\ref{figtestPDF}   for  one   realisation   of   the
$\mathrm{\Lambda}$CDM model with sources  at $z_s=1.4$.  The different
line styles  and color histograms show  the PDF when the  original map
has been  pixel-degraded by a  factor of $2$, $4$  and $8$ --  see the
figure caption for  more details.  It can be seen  that increasing the
pixel  size reduces  the number  of very  high and  very low  $\kappa$
pixels  as expected.   The impact  this has  on the  convergence power
spectrum is  shown in the  right hand panel  of Fig.~\ref{figtestPDF},
where the power spectra for different resolutions are plotted.  In the
bottom right hand panel are the relative residuals with respect to the
original, highest  resolution map.   These figures show  that reducing
the map resolution by a factor of two causes small differences of only
a few percent  in the power spectrum relative to  the original map for
$l \simlt 5\times 10^3$.  The discrepancies become larger for the maps
degraded  by  a factor  of  four  and  eight.   We conclude  that  our
calculations  are not  affected by  pixelation for  $l \simlt  5\times
10^3$ above few percent level.

It has  been shown in this  section that the evolution  of the density
power  spectrum in  different coupled  dark energy  models will  cause
significant changes to the convergence power spectrum as a function of
source redshift.  These differences  from $\Lambda$CDM can be positive
or negative  at a  particular source  redshift and  they can  be scale
dependent.  In the  following sections we will  investigate some other
statistics that  might be observationally  more practical in  terms of
direct measurements and noise estimations.

\subsubsection{Impact on the measured normalisation of the power spectrum for sources at 
different redshifts}

As noted  previously, the coupled  DM-DE models affect both  the power
spectrum  normalisation and  the small  scale non-linear  behaviour of
structure formation  when compared  with the $\Lambda$CDM  model.  One
way to  see this is  through the impact  on measurements of  the power
spectrum  normalisation.  The  power  spectrum  normalisation will  be
quantified by the standard $\sigma_8$  parameter which is the variance
of  the   mass  overdensity   density  within   a  sphere   of  radius
$8~h^{-1}$~Mpc at redshift zero with  only the linear evolution of the
power spectrum taken into account.   When $\sigma_8$ is measured using
data  at  redshifts larger  than  zero  a  correction factor  that  is
cosmology dependent must be applied  to translate the normalisation to
$z=0$.  If the underlying cosmology assumed in doing this procedure is
the  correct  one  then  $\sigma_8$  will not  depend  on  the  source
redshift.   In this  section we  investigate  what would  happen if  a
$\Lambda$CDM cosmology is assumed while the true cosmology corresponds
to one of our coupled dark  energy models.  This exercise will help us
gain some  insight on  when during cosmic  history these  models leave
most of their imprints on the lensing power spectrum.

\begin{figure*}
\includegraphics[width=0.33\hsize]{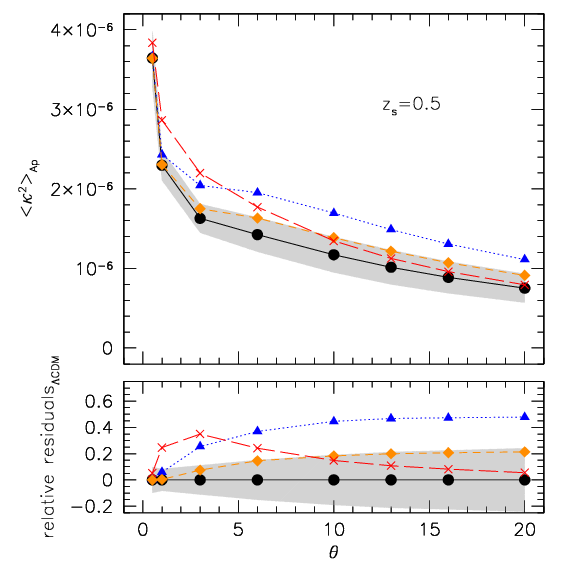}
\includegraphics[width=0.33\hsize]{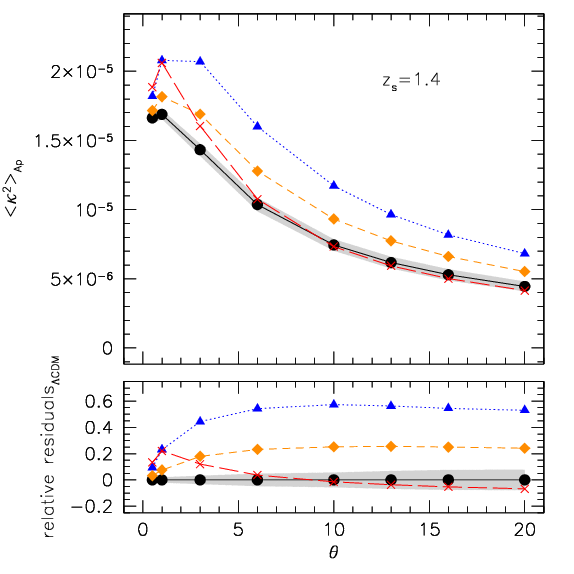}
 \includegraphics[width=0.33\hsize]{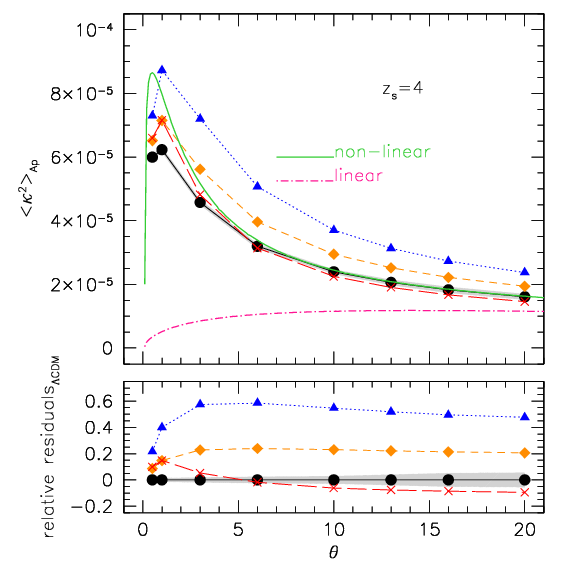}
 \caption{Top panels: variance of  the convergence field smoothed with
   a compensated aperture filter as  a function of the smoothing scale
   $\theta$, for sources at three different redshifts: $z_s=0.5,\,1.4$
   and $4$, from left to  right.  Bottom panels: relative residuals as
   a function of the smoothing  scale with respect to the measurements
   in  the  $\mathrm{\Lambda}$CDM  model.    The  grey  shaded  region
   represents  the  rms of  the  variance  computed on  the  different
   realizations of the $\mathrm{\Lambda}$CDM light-cone.  In the right
   panel, for sources at $z_s=4$, we also show predictions from linear
   and  non-linear  theory.   For   the  other  source  redshifts  the
   agreement is analogous to this one. \label{figkappa2Ap}}
\end{figure*}

The left hand panel of Fig.~\ref{figBP} shows the recovered $\sigma_8$
values as  a function of the  different source redshifts for  the four
cosmological models while  assuming $\mathrm{\Lambda}$CDM.  We measure
$\sigma_8$  integrating  the   non-linear  power  spectrum  \citep[see
  e.g.][]{camb,takahashi12}  up  to  the different  considered  source
redshifts assuming  the $\mathrm{\Lambda}$CDM  model; in this  way the
$\sigma_8$ value is directly related  to the measurement of the growth
factor associated  to a  $\mathrm{\Lambda}$CDM Universe.   Line styles
and colors are  the same as in Fig.~\ref{figpowerl},  the shaded light
grey and grey regions indicate $5$  and $10$ percent accuracies in the
measurement of $\sigma_8$  for the $\Lambda$CDM case.   Also shown are
the  true  $\sigma_8$  values  --  interpolated  using  linear  theory
\citep{baldi12}    --   for    each   model.     In   the    case   of
$\mathrm{\Lambda}$CDM the correct value is recovered, but in the other
models incorrect $\sigma_8$ values are  recovered and they change as a
function of source redshift.  Such  behaviour would therefore signal a
failure of  the underlying  $\Lambda $CDM  assumption.  In  EXP003 and
EXP008e3, $\sigma_8$ is underestimated, while  the opposite is true in
SUGRA003 for some of the source  redshift range.  These trends are due
to the  different evolution  of the  growth factor  and of  the Hubble
function in  the coupled DM-DE  models -- see Fig.   1 and Fig.   2 in
\cite{baldi12}.  It is interesting to  note that for the three coupled
models  we never  obtain  the value  of  $\sigma_8$ interpolated  from
linear theory at  the present time.  This means that  when fitting the
convergence power spectrum of the coupled DM-DE models both projection
effects and non-linearities contribute to the measured $\sigma_8$.

In  the right  hand panel  of  Fig.~\ref{figBP} we  show the  measured
convergence power spectrum in the different light-cone simulations for
sources at  $z_s=4$ -- solid,  dotted, dashed and dot-dashed  refer to
$\mathrm{\Lambda}$CDM,  EXP003, EXP008e3  and SUGRA003,  respectively.
For the three coupled DM-DE models we display best fitting theoretical
convergence  power  spectra   obtained  by  assuming  we   live  in  a
$\mathrm{\Lambda}$CDM universe.   In the  bottom panel we  present the
relative  deviation  of  the theoretical  convergence  power  spectrum
computed with  the best $\sigma_8$  and the one  linearly interpolated
from theory at $z=0$.  While for SUGRA003 model the relative deviation
is of the order of $1-2\%$,  for both EXP003 and EXP008e3 it manifests
larger values.

These  results  indicate that,  if  sufficient  accuracy is  attained,
measuring $\sigma_8$ from  shear maps as a function  of redshift would
lead to inconsistencies were one of these coupled dark energy models the
correct one.  To  observe these effects will require  large amounts of
very accurate  data of the kind  that will be provided  by future wide
surveys,  like the  Euclid  space  mission \citep{euclidredbook}.   An
additional  complication not  taken  into account  here  is that  weak
lensing studies are generally sensitive  to a combination of the power
spectrum normalisation,  $\sigma_8$, and the total  density of matter,
$\Omega_m$,              in               the              combination
$\sigma_8\left(\Omega_m/0.25\right)^{\alpha}$   where    $\alpha$   is
dependent on the weak lensing  statistic used: $\alpha = 0.46$, $0.53$
and $0.64$, for the shear two-point correlation function, the shear in
a  top-hat or  aperture mass,  respectively \citep[][see  also a  more
  extended   discussion   in  the   next   section]{fu08,kilbinger13}.
However,  $\Omega_m$  is  likely  to  be  well  constrained  by  other
observations such as the CMB.

\subsection{Other weak lensing statistics} 

Other statistics of  the convergence field besides  the power spectrum
may  help  to  probe  the   non-Gaussian  nature  of  the  probability
distribution  function  of   gravitational  lensing  observables.   In
particular, statistics like the variance and the skewness in a top-hat
or  compensated  filter  represent   interesting  tools  to  constrain
cosmological parameters and the dark energy evolution as a function of
redshift   \citep{fu08,kilbinger13,kitching14}.    In  this   context,
multi-plane  lensing simulations  are important  tools to  compute the
predicted variance  and skewness  of the  shear or  convergence field,
both  for  standard  and  non-standard  models.   In  particular,  the
skewness in  aperture is somehow  independent of the  normalization of
the power spectrum and represents  a strong indicator of the evolution
of $\Omega_m$ as a function of redshift \citep{schneider98}.

The interpretation  and modelling of  the lensing signals due  to large
scale structures  requires a  precise understanding of  the high-order
statistics of the lensing  field \citep{sato09}.  Several cosmological
analyses have been done by studying the mass aperture variance and the
skewness as  a function  of the smoothing  angle $\theta$,  adopting a
top-hat or  a compensated  aperture filter. These  measured quantities
have  the  advantage that  they  can  be  directly compared  with  the
theoretical predictions extracted from the convergence power spectrum.
However,  a good  model  of  the signal  on  small  scales, where  the
non-linear effects start to dominate,  is of fundamental importance to
interpreting observational data.  Since  the top-hat shear rms between
different  {\small  CoDECS} models,  but  only  at $z_s=1$,  has  been
studied by \citet{pace15}, we will present and discuss in this section
the  aperture-mass dispersion  and the  associated skewness  for three
different source redshifts.

The variance of  the filtered shear field as a  function of the filter
size  contains the  same information  as the  power spectrum,  but the
filter can  be made to  have compact support and  additional practical
advantages \citep{sato13}.  They can also have a different sensitivity
to the cosmological model and  be particularly dependent on non-linear
structure  formation.   In  addition, the  skewness  and  higher-order
statistics  can be  easily defined  and interpreted  for the  filtered
shear field.  We will investigate a particular choice of filter called
the compensated aperture filter:
\begin{equation}
Q_{\theta}(\vartheta) = \dfrac{6}{\pi \theta^2} \left( \dfrac{\vartheta}{\theta^2}\right) 
\left( 1 - \dfrac{\vartheta^2}{\theta^2} \right) \,
\end{equation}
 with  support $\vartheta=[0,\theta]$.  In Fourier space it is
\begin{equation}
W_{ap} = \dfrac{\sqrt{276}J_4(\xi)}{\xi^2}
\end{equation}
as   Fourier  counterpart   \citep{harnois-deraps12}.    It  has   the
attractive  property of  being  well-localized in  Fourier space  near
wavenumber $l \sim  5/\theta$.  We apply this filter  in Fourier space
with zero-padding to reduce boundary effects.  Since the average value
of the convergence on each plane is zero, the variance across the map,
typically indicated as $\langle M_{Ap}^2\rangle$, can be computed from
the convergence power spectrum performing the following integral:
\begin{equation}
  \langle k^2 \rangle_{Ap} =\langle M_{Ap}^2\rangle = \dfrac{1}{\left(2\pi\right)^2} \int \mathrm{d}l\,l P_{\kappa}(l)
  W^2_{Ap}(l \theta)\,.
\end{equation}
Similarly, it  is possible to  define the large scale  structure noise
\citep{hoekstra03}  adopting  a   different  compensated  filter  (see
Appendix \ref{apsec} for more discussion about this).

\begin{figure*}
\includegraphics[width=0.33\hsize]{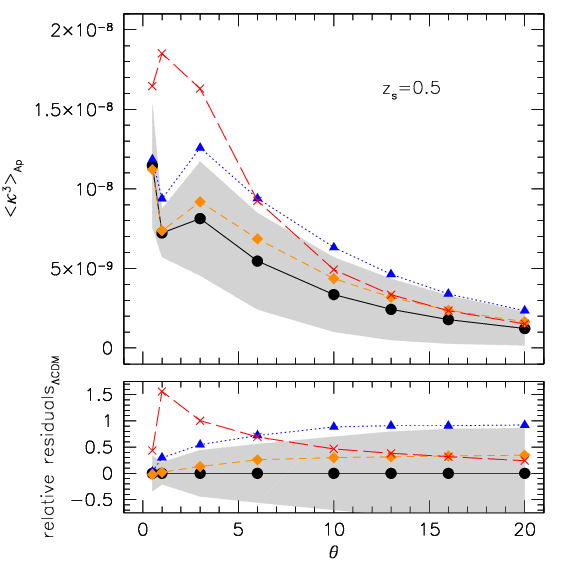}
\includegraphics[width=0.33\hsize]{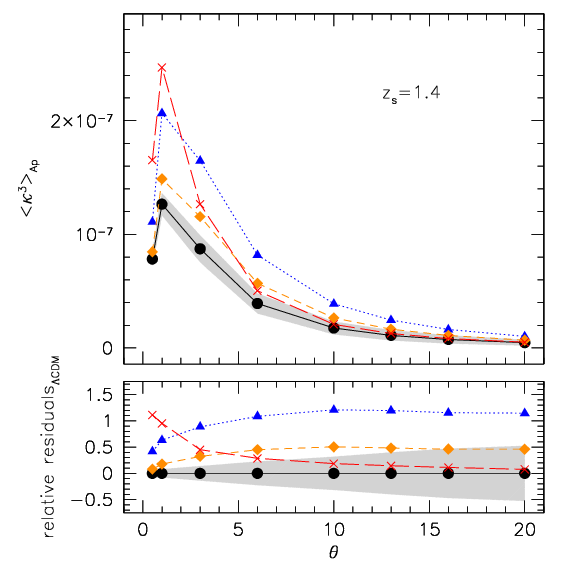}
\includegraphics[width=0.33\hsize]{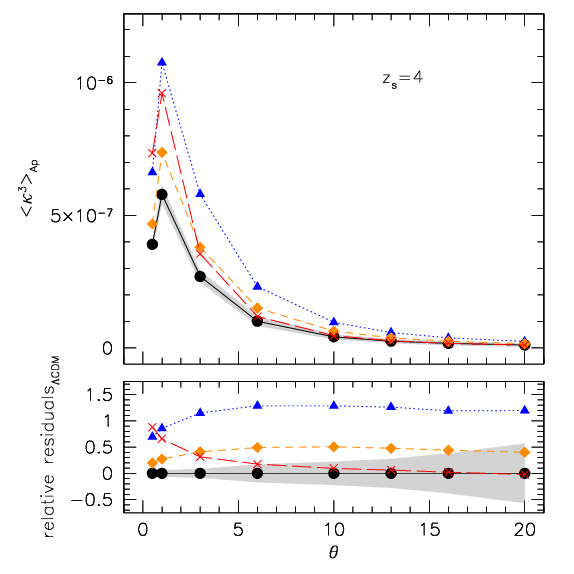}
\caption{As   Fig.~\ref{figkappa2Ap},   but   for  skewness   of   the
  convergence field.
  \label{figkappa3Ap}}
\end{figure*}

In  Fig.~\ref{figkappa2Ap}, we  show the  variance of  the convergence
field as a function of the  smoothing scale $\theta$ with this filter.
The  different line  styles and  colors refer  to the  four considered
cosmological models and the shaded grey region encloses the rms of the
measurement   performed  in   the   different   realizations  of   the
$\mathrm{\Lambda}$CDM light-cones.  The measurements are presented for
three source redshifts: $z_s=0.5$, 1.4  and 4.  The bottom panels show
the relative  residuals of the  measurements in the  different coupled
models with respect to the $\mathrm{\Lambda}$CDM ones.  We notice that
the  variance of  the measurement  in the  $\mathrm{\Lambda}$CDM model
decreases as a  function of the source redshift, as  expected from the
evolution of the clustering of the  dark matter in the Universe. While
at low  redshifts the Universe is  more clustered and so  we may trace
rays  through clusters  and voids  --  enhancing the  variance of  the
measurements in different realisations, at high redshifts the Universe
is more homogenous and so  the variance through different realisations
is expected to be smaller, as we confirmed in Fig.~\ref{figkappa2Ap}.

To demonstrate how much of this signal depends on non-linear structure
formation we show $\langle M^2_{Ap}  \rangle$ calculated with only the
linear    power    spectrum   in    the    right    hand   panel    of
Fig.~\ref{figkappa2Ap} for $z_s=4$ in the $\mathrm{\Lambda}$CDM model.
It can be seen that non-linear structure formation has a dominant role
below $\theta \simeq 20$ arcmin.

The  agreement  between  non-linear theoretical  predictions  and  our
simulations  for  $\mathrm{\Lambda}$CDM  are  quite  good  aside  from
limitations in the  simulations at small scales due  to numerical mass
and force  resolution limitations.   As with  the power  spectrum, the
deviations  from $\mathrm{\Lambda}$CDM  in Fig.~\ref{figkappa2Ap}  are
positive for EXP003 and EXP008e3 (by about 50 and 20 percent -- as for
the  convergence power  spectrum).   For  SUGRA003, $\langle  M^2_{Ap}
(\theta) \rangle$  is larger  than for $\mathrm{\Lambda}$CDM  at small
$\theta$ and becomes  smaller at large scales  mirroring the behaviour
in Fig.~\ref{figpowerl}.   This trend is  a consequence of  a slightly
lower  $\sigma_8$ governing  the  large $\theta$  behaviour while  for
small  scales the  boost in  variance  comes from  differences in  the
structure of halos: in SUGRA003  haloes are more concentrated and have
more  substructures  mainly  at  low  redshifts  \citep[as  was  shown
  e.g. by][]{giocoli13}.

The  filtered variance's  strong  dependence  on non-linear  structure
leads one to  think that the skewness of the  filtered convergence map
might be a good discriminator between models.  The skewness is plotted
in  Fig.~\ref{figkappa3Ap} as  a  function of  $\theta$,  for all  the
cosmological models, and once again for three source redshifts.  As in
the case of the variance, the  measurements in the EXP003 and EXP008e3
cosmologies are  larger than those in  $\mathrm{\Lambda}$CDM, but with
the relative  residuals that are a  factor of two larger  than for the
variance.   In  addition,  for  large angles  the  relative  residuals
between $\mathrm{\Lambda}$CDM and SUGRA003 are very small but increase
for small $\theta$,  a signature of the higher  small scale clustering
present  in this  case, the  high  concentration of  haloes and  their
clumpiness.  For a  fixed angular scale the  difference becomes larger
for  smaller source  redshift  owing  to the  build  up of  non-linear
structure  at late  times  in the  SUGRA003  model.  These  difference
between models are more evident here than for the variance.

\begin{figure*}
\includegraphics[width=0.49\hsize]{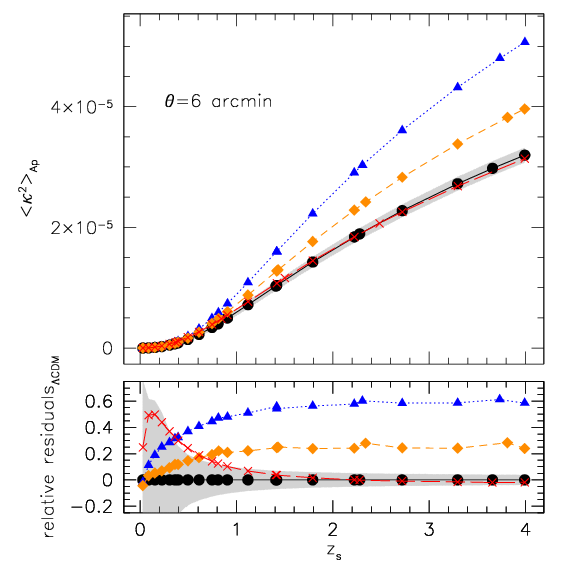}
\includegraphics[width=0.49\hsize]{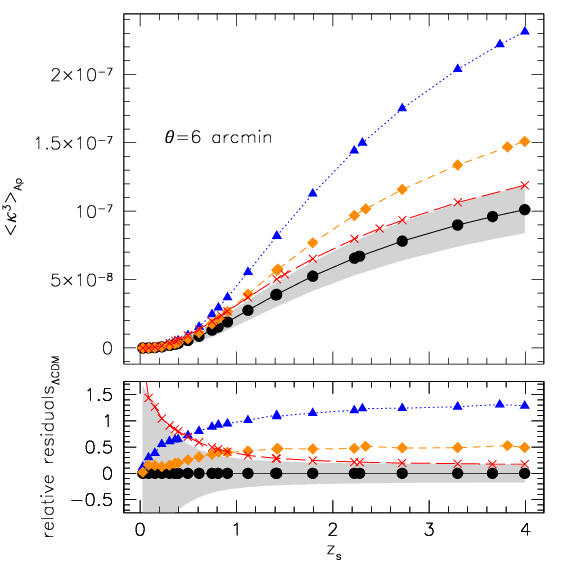}
\caption{Variance  (left  panel) and  skewness  (right  panel) of  the
  convergence field  smoothed with a compensated  aperture filter with
  $\theta=6$ arcmin as a function of the source redshifts.  The bottom
  panels show the relative residuals  of the variance and the skewness
  with respect to the  $\Lambda \mathrm{CDM}$ measurements. The shaded
  grey  region represents  the  rms  of the  moments  computed on  the
  different    realizations    of     the    $\Lambda    \mathrm{CDM}$
  light-cones.\label{k2k3Apz}}
\end{figure*}

In Fig.~\ref{k2k3Apz}  we show  the variance  (left) and  the skewness
(right) of the convergence field as  a function of the source redshift
while fixing the  scale of the filter to 6  arcmin which is comparable
to  the typical  scale of  the central  region of  galaxy clusters  at
intermediate redshifts.  The bottom panels again show the residuals of
the  measurements with  respect  to  $\mathrm{\Lambda}$CDM.  Both  the
variance and the  skewness grow as a function of  the source redshifts
because of  the increased  path lengths  and the  additional structure
along  the paths.   For $z_s>0.5$  all the  models tend  to present  a
constant  bias  with  respect to  $\mathrm{\Lambda}$CDM:  positive  in
EXP003  and EXP008e3  while  almost vanishing  in  the bouncing  model
SUGRA003.  Note that the relative difference in the skewness is almost
double  than in  the variance.   While the  behaviour of  the bouncing
model  SUGRA003  is  within   the  rms  of  the  $\mathrm{\Lambda}$CDM
measurements,  the  EXP003 and  EXP008e3  are  quite distinct  with  a
positive bias of about $60$ percent  and $20$ percent for the variance
and $120$ percent and $50$ percent for the skewness for $z_s>0.5$.

The behaviour of the variance and skewness of the filtered convergence
field  highlights  the  possibility  of  using  sources  at  different
redshifts  and  with different  smoothing  scales  to investigate  the
coupling between the dark components of the universe.

\subsection{Probability distribution function from source redshift
  distributions}

\begin{figure*}
\includegraphics[width=0.45\hsize]{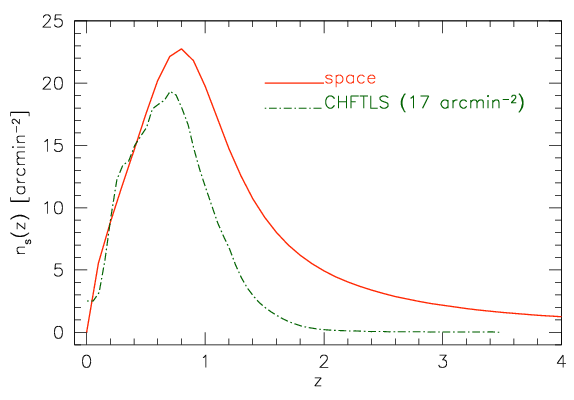}
\includegraphics[width=0.45\hsize]{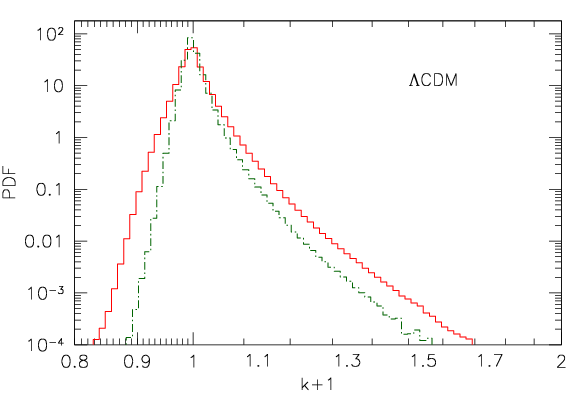}
\includegraphics[width=0.45\hsize]{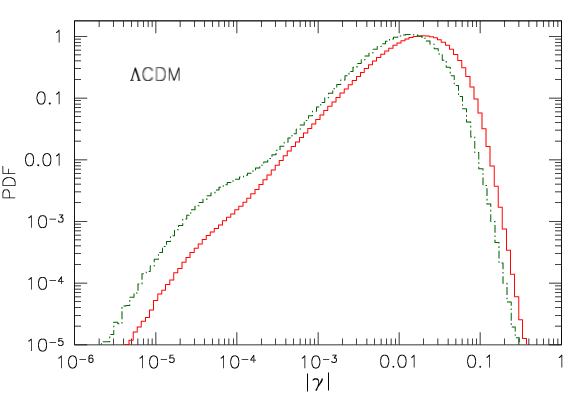}
\includegraphics[width=0.45\hsize]{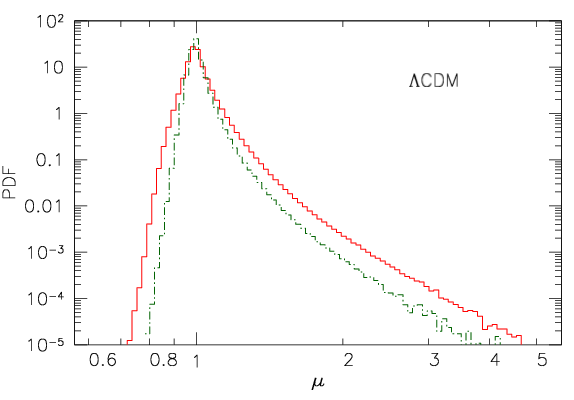}
\caption{Top left:  source redshift  distribution expected  for space-
  and ground-based  observations.  Top  right, bottom left  and bottom
  right PDF of convergence, shear and magnification extracted from the
  twenty-five $\mathrm{\Lambda}$CDM light-cones randomly sampled eight
  times  with the  corresponding  source  redshift distribution.   The
  various  color and  line  style histograms  refer  to the  different
  considered cosmologies  (see Fig.~\ref{figpowerl}), the  shaded grey
  region  encloses  the  rms  of  the  measurement  performed  in  the
  different  light-cone  realizations   in  the  $\mathrm{\Lambda}$CDM
  model. \label{LCDMpdffig}}
\end{figure*}

In previous  sections we have  discussed the results for  the variance
and  the skewness  of the  convergence  field for  sources located  at
different  fixed  redshifts.  In  this  last  section we  explore  the
one-point   distribution   function    of   convergence,   shear   and
magnification  given instead  a source  redshift distribution.   We do
this in order to understand whether the simple one-point statistics of
the lensing field, and which  source redshift distribution can help us
to understand the  non-Gaussian nature of the fields  in the different
cosmological models.  In our analysis we consider two different source
redshift distributions  that typically correspond  to a ground-  and a
space-based weak lensing  survey.  For the ground-based  case we adopt
the  source  density  as  a   function  of  redshift  as  computed  by
\cite{kilbinger13} --  the data points  are publicly available  on the
CFHTLenS
webpage\footnote{\href{http://www.cfhtlens.org/astronomers/cosmological-data-products}{http://www.cfhtlens.org/astronomers/cosmological-data-products}}
-- while for  the space-based we consider  the parametrization adopted
by  \citet{boldrin12}  which has  been  extracted  from a  Euclid-like
observation of  the Hubble  Ultra Deep Field  performed with  the code
\texttt{SkyLens}  \citep{meneghetti08}.   In  the top  left  panel  of
Fig.~\ref{LCDMpdffig}  we   show  the  two  adopted   source  redshift
distribution normalizing  the CFHTLS  one to have  a total  density of
sources of  about $17$ galaxies  per arcmin$^2$. What is  most evident
about  the two  is  that the  high redshift  tail  in the  space-based
observation extends to higher redshifts while in the ground-based case
there are  almost no galaxies  above $z=2$.   In order to  extract the
convergence and  shear catalogues  from one light-cone  realization we
proceed in the following way: (i) given a source redshift distribution
we  compute the  number of  expected  sources in  a redshift  interval
$\mathrm{d}z$ that corresponds to  the difference in comoving distance
between two  source planes;  (ii) for  each $\mathrm{d}z$  we randomly
assign to each  source both a redshift and an  angular position in the
field of  view.  We  do not  take into account  any clustering  of the
sources  which   should  no  be   important  for  our   purpose.   The
corresponding lensing quantity (convergence or shear) is then linearly
interpolated in  redshift considering the values  computed between two
consecutive  planes  at  the  corresponding angular  position  of  the
source.    The  other   panels  of   Fig.~\ref{LCDMpdffig}  show   the
convergence, shear  and weak  lensing magnification for  the different
$\mathrm{\Lambda}$CDM light-cone  realizations, each  randomly sampled
twenty-five times  with the  according $n(z_s)$.   From the  figure we
notice that because the two adopted source redshift distributions have
different shapes, mainly  in the intermediate and  high redshift tail,
the one-point  lensing statistics from the  same simulated light-cones
tend  to be  different. While  the shape  of the  convergence and  the
magnification  are   broadened  going  from  ground-   to  space-based
observations, the shear distribution is shifted toward larger values.

\begin{figure*}
\includegraphics[width=0.33\hsize]{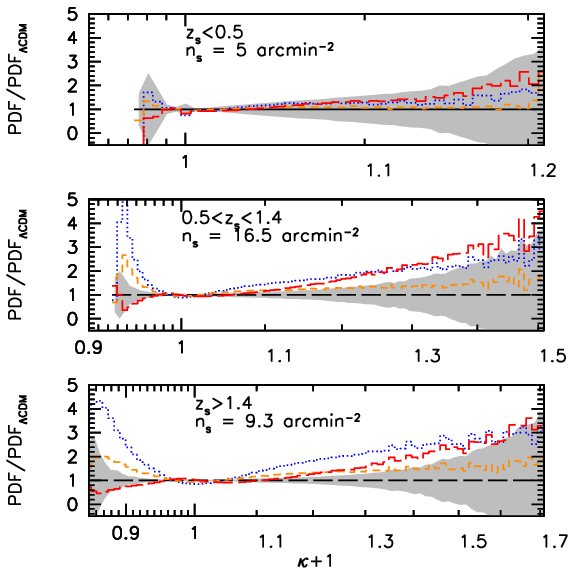}
\includegraphics[width=0.33\hsize]{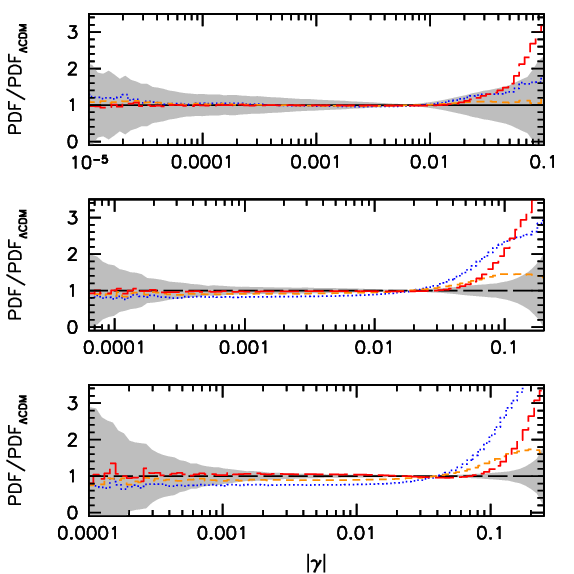}
\includegraphics[width=0.33\hsize]{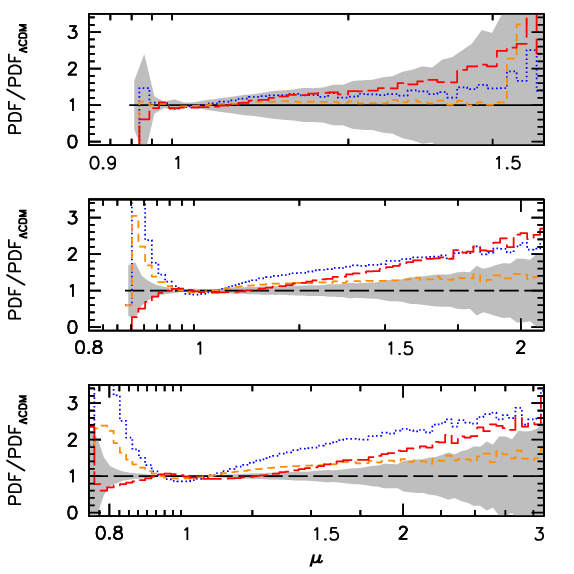}
\caption{PDFs of convergence (left column), shear (central column) and
  magnification (right  column) extracted from the  light-cones of the
  different    cosmologies    rescaled    with    respect    to    the
  $\mathrm{\Lambda}$CDM one,  assuming a  space based  source redshift
  distribution.  Rows  refer to  different  intervals  for the  source
  redshifts:  $z_s<0.5$,  $0.5<z_s<1.4$  and $z_s>1.4$,  from  top  to
  bottom.  Each light-cone  has been randomly sampled  eight time with
  the according source redshift distribution.\label{Eupdffig}}
\end{figure*}

In  Fig.~\ref{Eupdffig} we  show the  expected convergence,  shear and
magnification probability distributions assuming  a space based source
redshift distribution for the  four cosmologies, rescaled with respect
to the $\mathrm{\Lambda}$CDM one.   In comparison to \citet{pace15} we
notice that our  distributions are realistically more  extended in the
high-value tails  because of  the different way  adopted in  doing the
ray-tracing,  and of  the larger  resolution with  which the  maps are
resolved. While in  Fig.  6 by \citet{pace15}  the one-point statistic
manifests  itself  mainly  in  the different  initial  power  spectrum
normalisation  of the  coupled  models, in  Fig.~\ref{Eupdffig} it  is
possibile to  observe also a  more pronounced distinction in  the high
value tails. Particularly, in both the  PDF of the convergence and the
magnification,  the   small  scale   clustering  and  the   high  halo
concentration  in SUGRA003  model  raise the  high value  distribution
tails.  This  is also  evident in the  shear distributions  where both
SUGRA003   and   EXP003   are   well    outside   the   rms   of   the
$\mathrm{\Lambda}$CDM distribution for  $\gamma>0.075$.  The situation
is  different  for  the  corresponding PDF  extracted  from  the  same
light-cones considering  a ground-based source  redshift distribution,
not shown  here.  First  we notice  that the  source densities  in the
intermediate and  the high  redshift bins reduce  to $11.2$  and $0.8$
$\mathrm{arcmin}^{-2}$, respectively,  and that the SUGRA003  model is
enclosed within the rms of the $\Lambda$CDM for large values.  In this
case  only EXP003  appears  distinguishable from  $\Lambda $CDM.   The
signature of enhanced high redshift growth in the considered models is
much less distinguishable with the ground-based lensing survey than it
is with a space based survey, as also discussed in \citet{beynon12}.

\begin{figure}
\includegraphics[width=\hsize]{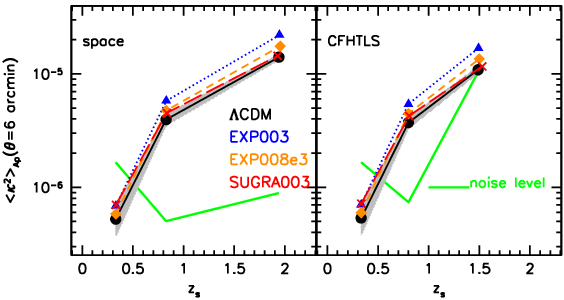}
\caption{Variance  of  the  convergence   field  computed  within  the
  different light-cone  realisations and cosmologies adopting  a space
  and a ground-based (CFHTLS)  source redshift distribution. Different
  data points and  colors are the same as in  Fig. \ref{k2k3Apz}.  The
  solid  (green)   line  represents  the  corresponding   noise  level
  associated to the intrinsic  galaxy ellipticity distribution and the
  source number  density. This noise level  is for a 25  square degree
  field. \label{figBEN}}
\end{figure}

In Fig.~\ref{figBEN} we display the  variance of the convergence field
computed from different light-cone realisations in three redshift bins
for a space (left) and  ground-based (right), assuming a CFHTLS source
redshift distribution. Black, blue, orange and red colors refer to the
$\Lambda$CDM, EXP003, EXP008e3 and  SUGRA003 models, respectively.  In
each panel we also show with a green solid line the noise level in the
three redshift bins related  to the intrinsic ellipticity distribution
$\sigma_{\epsilon}=0.25$  and  to   the  corresponding  source  number
density     $n_{g}$    contained     in     the    aperture     filter
\citep{schneider98,vanwaerbeke00}.  This is the  noise for a 25 square
degree field.   Neglecting systematic errors,  the noise for  a larger
survey will go down by roughly a factor of one over the square root of
the survey area.   From this plot it  can be seen that  a ground based
survey  will have  difficulty distinguishing  between the  models, but
that a  survey like  Euclid would be  expected to  clearly distinguish
between them.

\section{Summary and Conclusions}
\label{secsummary}
Several studies  have been  conducted to understand  which observables
are  most  suitable to  investigate  coupled  Dark Energy  cosmologies
\citep{beynon12,giocoli13,carbone13,pace15}.   Our  aim  here  was  to
extend the ray-tracing analyses  of \cite{carbone13} and \cite{pace15}
by  investigating  whether a  tomographic  slicing  of the  background
sources  within high  resolution simulated  light-cones might  provide
additional information to distinguish  coupled dark energy models from
the standard  $\Lambda $CDM cosmology  and possibly the  different cDE
models  from  each other.   Our  main  results  can be  summarized  as
following:
\begin{itemize}
\item While the  convergence power spectra in the  EXP003 and EXP008e3
  models  present only  a  higher normalisation  with  respect to  the
  $\mathrm{\Lambda}$CDM  model, the  power  spectrum  in the  SUGRA003
  model   exhibits   a   more    complex   behaviour.    Compared   to
  $\mathrm{\Lambda}$CDM, in this  model structure formation progresses
  rapidly at high  redshift and slows down at  low redshift, resulting
  in a comparable value of  $\sigma_8$ at $z=0$ once the normalisation
  is  fixed  using  the  CMB.    The  large  scale  behaviour  of  the
  convergence  power  spectrum  reflects  the  slower  growth  at  low
  redshifts and it  is weakly suppressed.  This suppression  is due to
  the change  of sign of  the drag term in  this class of  cDE models,
  which are  characterised by  a ``bounce" of  the Dark  Energy scalar
  field  whose motion  changes direction  \citep[see ][for  a detailed
    discussion of Bouncing Coupled  Dark Energy]{Baldi_2011c}.  In the
  nonlinear regime this  effect determines a faster  collapse of bound
  structures that results  in a higher average  concentration of halos
  and  in a  larger abundance  of substructures  \citep[see e.g.   the
    discussion in][]{giocoli13}.  This is  reflected in an enhancement
  of the convergence  power spectrum at small angular  scales -- large
  $l$.
\item The coupling between the dark components can also be seen in the
  lensing signals as  a function of source  redshifts.  In particular,
  if  $\sigma_8$  is measured  from  the  convergence power  spectrum,
  assuming $\mathrm{\Lambda}$CDM and fixing $\Omega_m$ in the fitting,
  the  result will  change as  a function  of source  redshift in  cDE
  cosmologies,    i.e.     they    will    be     inconsistent    with
  $\mathrm{\Lambda}$CDM.
  \item The  cross-correlation between  the convergence  for different
    source redshifts  is significantly enhanced in  the SUGRA003 model
    while it  is slightly  suppressed in the  EXP003 and  the EXP008e3
    models relative to what is expected in $\mathrm{\Lambda}$CDM, this
    because  of  the  high  concentration and  the  high  small  scale
    clustering  that  manifest as  consequence  of  the high  and  low
    redshift enhanced and suppressed growth rate that characterize the
    SUGRA003 model.
  \item  The  aperture  mass  statistic also  exhibits  signs  of  the
    coupling between dark  matter and dark energy.   In particular for
    $\theta>5$ arcmin  and sources $z_s>1$ EXP003  and EXP008e3 differ
    by about $50$ percent and  $20$ percent, respectively, as compared
    to $\mathrm{\Lambda}$CDM.  The skewness of a compensated aperture,
    measuring the  non-Gaussian nature of the  convergence field, also
    reflects differences between the  various models, specifically the
    higher level  of dark matter  clustering in the SUGRA003  model is
    evident for small filtering scales.
  \item  The different  models analyzed  in  the work  also result  in
    distinct  PDF of  the  lensing signals  -  convergence, shear  and
    magnification  -  extracted from  a  space  based source  redshift
    distribution.  The various models not only manifest differences in
    the intermediate values, as  discussed by \citet{pace15}, but also
    in the high value tails.
\end{itemize}

Therefore, it  emerges from the  analyses performed in this  work that
the differences between the coupled dark matter-dark energy models and
the standard CDM  can be explained not only in  terms of the different
normalizations of the  linear matter power spectrum but  also in terms
of  the  distinct halo  properties  and  small scale  clustering  that
characterize the dark matter  component.  These manifest themselves in
different weak-lensing observables at small angular scales and also in
the one  point statistics  taken in different  redshift bins,  given a
space  based  source  redshift   distribution  that  extends  to  high
redshifts.  In particular, in this last case, both SUGRA003 and EXP003
models  appear to  be  easily distinguishable  from  $\Lambda $CDM  by
future weak lensing surveys.

\section*{Acknowledgements}
We thank the  anonymous referee for her/his useful  comments that help
to improve  the presentation of our  results.  CG thanks CNES  for its
financial  support.  CG  and RBM's  research  is part  of the  project
GLENCO, funded under the  European Seventh Framework Programme, Ideas,
Grant  Agreement n.   259349.  MB  acknowledges support  by the  Marie
Curie  Intra European  Fellowship  ``SIDUN" within  the 7th  Framework
Programme  of  the  European  Commission.   We  acknowledge  financial
contributions from contracts ASI/INAF I/023/12/0  and by the PRIN MIUR
2010-2011 ``The  dark Universe  and the  cosmic evolution  of baryons:
from current surveys to Euclid''.  CG, MB, and LM also acknowledge the
financial contribution  by the  PRIN INAF 2012  ``The Universe  in the
box:  multiscale simulations  of  cosmic structure''.   We also  thank
Cosimo Fedeli, Federico Marulli,  Michele Moresco and Mauro Roncarelli
for useful discussions.  GC thank  Giulia Despali, Giuseppe Tormen and
Vincenzo  Mezzalira to  have host  part of  the computer  jobs run  to
perform the ray-tracing.

\appendix

\section{Large scale structure noise in cluster shear profiles}
\label{apsec}

The results  obtained in this paper  can also be used  to evaluate how
the  level of  noise  produced by  the large  scale  structure on  the
estimates   of  galaxy   cluster   masses  from   weak  lensing   data
\citep{bahe12,hoekstra12,merten14,giocoli14,sereno14,vonderlinden14a,vonderlinden14b}
are dependent on the cosmological background.

Several algorithms have been developed to invert the lens equation and
derive     the     cluster's    projected     mass     \citep{jullo07,
  merten09,zitrin11a}.    Since   galaxy   clusters  have   a   matter
distribution  that on  average  can  be described  by  a well  defined
density profile  \citep{navarro96,navarro04}, a  direct way  to weight
them  is  to  fit  the (spherically  binned)  measured  shear  profile
adopting      the       corresponding      theoretical      prediction
\citep{bartelmann96,giocoli12}.

However,  the light  traveling from  sources located  behind a  galaxy
cluster is deflected not only by cluster matter distribution, but also
by  all  the   matter  it  encounters  along   its  trajectory.   This
uncorrelated matter  density distribution  contributes to  the lensing
signal,       and      affects       the      shear       measurements
\citep{schneider98,hoekstra03,hoekstra11}.

This lensing  noise, which depends on  the source redshift and  on the
angular  size $\theta$  of  the  annulus we  are  looking  at, can  be
computed analytically from the convergence power spectrum adopting the
formalism developed by \citet{hoekstra03}:
\begin{equation}
\sigma^2_{LSS}(\theta) = \dfrac{1}{2 \pi} \int \mathrm{d} l \, l
P_{\kappa}(l) J^2_2(l \theta)\ ,
\end{equation} 
where  $J_2$ represents  the  second-order Bessel  function  and is  a
particular  choice of  the aperture  mass statistic  averaged over  an
annulus ranging  from ($\theta - \delta  \theta/2$) to ($\theta+\delta
\theta/2$).

\begin{figure*}
\includegraphics[width=0.33\hsize]{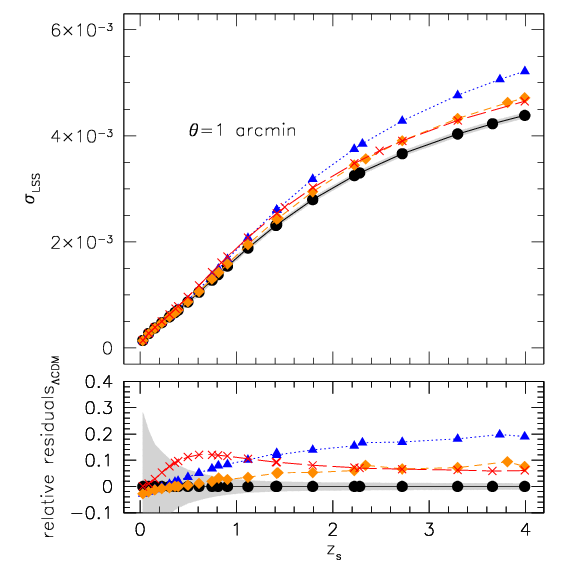}
\includegraphics[width=0.33\hsize]{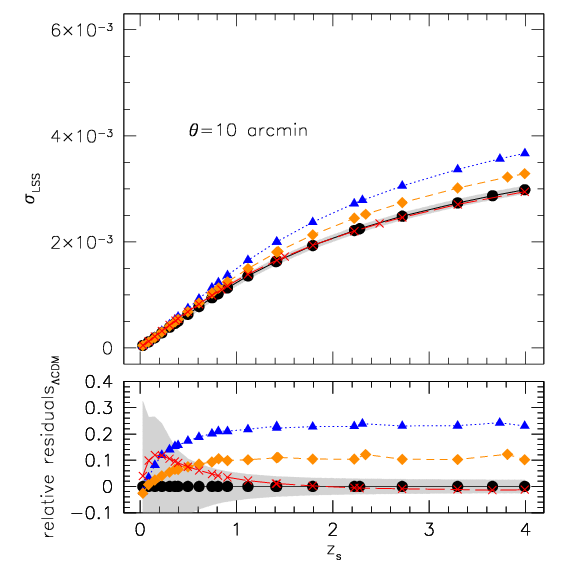}
\includegraphics[width=0.33\hsize]{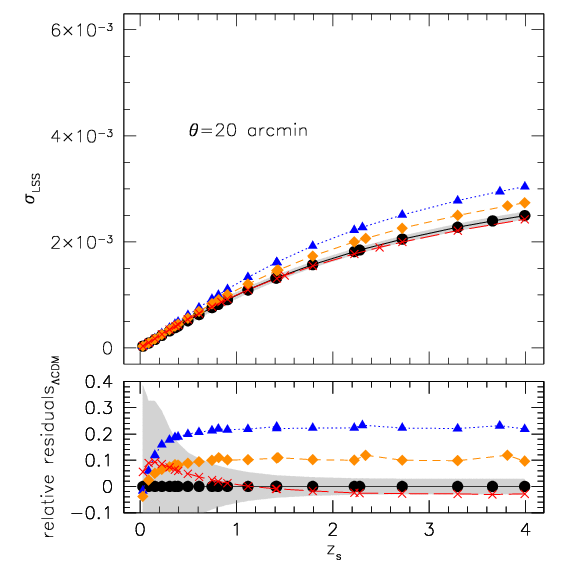}
\caption{Large scale  structure noise  computed adopting  a particular
  choice  of the  aperture mass  statistic, averaged  over an  annulus
  ranging     from     ($\theta     -     \delta     \theta/2$)     to
  ($\theta+\delta\theta/2$),  as a  function  of  the source  redshift
  measured from the  convergence maps extracted of  the light-cones of
  the four considered {\small  CoDECS} simulations.  Left, central and
  right panel refer  to different angular scales  $\theta$; colors and
  line styles are as in Fig.~\ref{figkappa2Ap}. \label{fsignal}}
\end{figure*}

In Fig.~\ref{fsignal}  we show,  for the four  considered cosmologies,
the noise produced by the  LSS on spherically averaged shear profiles.
The results are presented for three typical scales ($\theta$=1, 10 and
20 arcmin) and considering sources up to redshift $z_s=4$.  As already
done in  the case  of the  aperture mass  statistics, the  results are
directly extracted from  the convergence maps that  have been smoothed
using the appropriate $J_2$ kernel.

From  the figure  we notice  that the  noise produced  by large  scale
structures  is  higher   in  the  cDE  models  with   respect  to  the
$\mathrm{\Lambda}$CDM  cosmology. This  is particularly  true for  the
EXP003 and  EXP008e3 models:  for $\theta>10$  arcmin and  for sources
with $z_s>0.5$  they present values that  are about 20 and  10 percent
larger  than  the one  measured  in  the $\mathrm{\Lambda}$CDM  model,
respectively.  For the SUGRA003 model,  a smaller increase of noise is
evident  only  for  $\theta=  1$  arcmin and/or  for  sources  at  low
redshift.  {\color{red}  We  underline  also   that  the  rms  of  the
  measurement  $\mathrm{\Lambda}$CDM   decreases  as  a   function  of
  redshifts  because of  a  combination of  ($i$)  the choice  filter,
  ($ii$) the field of view and ($iii$) the clustering of the haloes as
  a function  of redshifts --  at high  redshift the universe  is more
  homogenius.}

\bibliographystyle{mn2e}
\bibliography{lensingTomographyCoDECS.bbl}
\label{lastpage}
\end{document}